\documentclass[]{aastex631}

\newcommand{\ang}{\text{\AA}}
\newcommand{\lya}{Ly$\alpha$}

\begin{document}

\title{Investigating the High-energy Radiation Environment of Planets in Sun-like Binary Systems}

\author[0000-0003-1369-8551]{Patrick R. Behr}
\affiliation{Department of Astrophysical and Planetary Sciences, University of Colorado, Boulder\\
Boulder, CO, 80309, USA}
\affiliation{Laboratory for Atmospheric and Space Physics\\
1234 Innovation Dr, Boulder, CO, 80303, USA}

\author[0000-0002-1002-3674]{Kevin France}
\affiliation{Department of Astrophysical and Planetary Sciences, University of Colorado, Boulder\\
Boulder, CO, 80309, USA}
\affiliation{Laboratory for Atmospheric and Space Physics\\
1234 Innovation Dr, Boulder, CO, 80303, USA}

\author{Nicholas Kruczek}
\affiliation{Laboratory for Atmospheric and Space Physics\\
1234 Innovation Dr, Boulder, CO, 80303, USA}

\author[0000-0001-7131-7978]{Nicholas Nell}
\affiliation{Laboratory for Atmospheric and Space Physics\\
1234 Innovation Dr, Boulder, CO, 80303, USA}

\author[0000-0002-2129-0292]{Brian Fleming}
\affiliation{Department of Astrophysical and Planetary Sciences, University of Colorado, Boulder\\
Boulder, CO, 80309, USA}
\affiliation{Laboratory for Atmospheric and Space Physics\\
1234 Innovation Dr, Boulder, CO, 80303, USA}

\author{Stefan Ulrich}
\affiliation{Laboratory for Atmospheric and Space Physics\\
1234 Innovation Dr, Boulder, CO, 80303, USA}

\author[0000-0002-7119-2543]{Girish M. Duvvuri}
\affiliation{Vanderbilt University\\
Nashville, TN, USA}

\author[0000-0002-3191-2200]{Amy Louca}
\affiliation{Leiden University\\
Leiden, Netherlands}

\author[0000-0002-3191-2200]{Yamila Miguel}
\affiliation{Leiden University\\
Leiden, Netherlands}

\begin{abstract}
Far-ultraviolet (FUV) radiation is a driving source of photochemistry in planetary atmospheres. Proper interpretation of atmospheric observations requires a full understanding of the radiation environment that a planet is exposed to. Using the Suborbital Imaging Spectrograph for Transition-region Irradiance from Nearby Exoplanet host stars (SISTINE) rocket-borne spectrograph, we observed the Sun-like binary system $\alpha$ Centauri AB and captured the FUV spectrum of both stars simultaneously. Our spectra cover 980--1570 \ang\, providing the broadest FUV wavelength coverage taken in a single exposure and spanning several key stellar emission features which are important photochemical drivers. Combining the SISTINE spectrum with archival observations, model spectra, and a novel stellar activity model, we have created spectral energy distributions (SEDs) spanning 5 \ang--1 mm for both $\alpha$ Centauri A and B. We use the SEDs to estimate the total high-energy flux (X-ray--UV) incident on a hypothetical exoplanet orbiting $\alpha$ Centauri A. Because the incident flux varies over time due to the orbit of the stellar companion and the activity level of each star, we use the \texttt{VULCAN} photochemical kinetics code to estimate atmospheric chemical abundances in the case of minimum and maximum flux exposure. Our results indicate that enhanced atmospheric mass loss due to stellar binarity will likely not be an issue for future exoplanet-hunting missions such as the Habitable Worlds Observatory when searching for Earth-like planets around Sun-like stars.

\end{abstract}


\section{Introduction} \label{sec:intro}
High energy stellar radiation is a driver of photochemistry in the upper atmosphere of exoplanets. Of particular importance are hydrogen, carbon, nitrogen, and oxygen bearing molecules. These molecules---such as H$_2$O, CO, CO$_2$, CH$_4$, N$_2$, and NH$_3$---are abundant in exoplanetary atmospheres and are readily destroyed by UV photons due to their large photodissociation cross sections at wavelengths less than $\sim$2200 \ang\ \citep{fortney_hot_2021,madhusudhan_exoplanetary_2019,loyd_muscles_2016}. When these molecules are dissociated by UV photons, the constituent atomic species can combine to form new molecules or, particularly in the case of H, escape the atmosphere entirely, leading to atmospheric mass loss \citep{vidal-madjar_detection_2004,murray-clay_atmospheric_2009,hu_photochemistry_2012}. The destruction and recombination of C and H bearing molecules also influences the abundance of hydrocarbon hazes. These hazes are detectable in transmission spectra of exoplanet atmospheres as a flattening the short wavelength end of the spectra, potentially obscuring absorption features in those regions \citep{he_laboratory_2018,kawashima_theoretical_2018}.

H I Lyman-$\alpha$ is of particular interest because it dominates the photodissociation of H$_2$O and CH$_4$ \citep{miguel_effect_2015} and is the brightest FUV emission line in the spectra of almost all cool stars \citep{france_ultraviolet_2013,france_2016_muscles,youngblood_2016_muscles}. Photolysis of H$_2$O is a major source of atomic H and OH radicals which are necessary in many photochemical reaction chains. Recent observations of the exoplanet WASP-39b with the James Webb Space Telescope (JWST) showed absorption features of SO$_2$ in the planet's atmosphere; this is the first direct evidence of a photochemically produced species in an exoplanetary atmosphere, and the first step in the chemical reaction chain is the destruction of water molecules \citep{tsai_photochemically_2023}. Other notable stellar FUV emission lines are \ion{O}{6} ($\lambda \lambda1032,1038$ \ang), \ion{Si}{3} ($\lambda1206$ \ang), \ion{N}{5} ($\lambda \lambda1238,1243$ \ang), \ion{C}{2} ($\lambda\lambda1334,1334$ \ang), \ion{Si}{4} ($\lambda\lambda1394,1403$ \ang), and \ion{C}{4} ($\lambda \lambda1548,1551$ \ang), as well as the tail of the photospheric and chromospheric blackbody emission for stars with effective temperatures similar to or hotter than the Sun. These strong FUV emission lines can generate O$_3$ via photolysis of CO$_2$ and O$_2$. O$_3$ can then be destroyed by near-ultraviolet (NUV; 2000-4000 Å) photons and the resulting atomic O may recombine into O$_2$; however, if the stellar FUV/NUV flux ratio is large or small, these processes may favor the generation of one molecule over the other and an abiotic buildup of O$_3$ and/or O$_2$ is possible \citep{meadows_exoplanet_2018,schwieterman_overview_2024,ranjan_importance_2023}. Oxygen and ozone have the potential to serve as biosignatures in exoplanetary atmospheres and thus it is important to understand the stellar UV radiation environment in order to distinguish between oxygen buildup due to biotic processes, such as oxygenic photosynthesis, and buildup due to abiotic processes.

While FUV and NUV radiation drive photochemical processes, shorter wavelength photons in the extreme-ultraviolet (EUV; 100-912 Å) can cause an atmosphere to escape from the planet altogether. EUV photons heat the upper atmosphere to extreme temperatures, driving rapid atmospheric escape. The high flow rate of escaping H can then drag along heavier elements such as O, C, Fe, and Mg in a hydrodynamic flow \citep{vidal-madjar_detection_2004,sing_hubble_2019}. However, stellar EUV radiation is heavily attenuated by H and He in the interstellar medium (ISM), making it difficult to observe the EUV spectrum for most stars. The Extreme Ultraviolet Explorer \citep[EUVE;][]{bowyer_1991_euve} was capable of observing the EUV spectrum of some nearby stars which were not completely attenuated but since its decommission in 2001, there is currently no operating facility capable of measuring stellar EUV emission. Options for estimating the EUV emission include measurements of the intrinsic Ly$\alpha$ and other transition-region emission line fluxes, which allows for coarse estimates of the EUV flux \citep{linsky_intrinsic_2014,france_2018_fuv,france_2025_euv} or, when combined with measurements of other chromospheric, transition-region, and coronal emission lines, differential emission measure techniques (DEMs) can create model spectra by fitting the temperature profile of the stellar atmosphere \citep{warren_1998_dem,sanz-forcada_2011_dem,duvvuri_reconstructing_2021}.

Currently, a major goal of the upcoming Habitable Worlds Observatory (HWO) is to perform an exo-Earth survey capable of characterizing $\sim25$ Earth-like planets within the habitable zone of their stars \citep{arney_2025_hwo}. In preparation for this survey, significant effort has been made in characterizing the high-energy emission of Sun-like FGK stars. The MUSCLES survey was extended to include SEDs of 8 additional FGK exoplanet-hosting stars, the majority of which were comparable to the Sun in their X-ray-to-bolometric luminosity ratios ($L_\mathrm{X}/L_\mathrm{bol}$) and UV activity levels and less active than comparable field stars \citep{behr_muscles_2023}. \citet{binder_2024_hwo} performed an analysis of XMM-Newton and Chandra X-ray observations of 57 nearby stars, of which 27 FGK stars exhibit $L_\mathrm{X}/L_\mathrm{bol}$ similar to the Sun. Studies of the high-energy evolution of the Sun and Sun-like stars have shown that the amount of X-ray and EUV radiation is highly dependent on the age of the star (with young stars having higher fluxes) as well as initial stellar rotation rate \citep{johnstone_2021_xuv,france_2025_euv}. This should be kept in mind when choosing target systems, as current lists of potential candidates \citep[e.g.,][]{mamajek_2024_exep-list,binder_2024_hwo} contain stars ranging in age from a few 100 Myr to $>10$ Gyr.

A complicating factor in the search for habitable planets is stellar activity, particularly for low-mass stars which have frequent flaring events that dramatically change the FUV emission on timescales of minutes \citep{loyd_muscles_2018,froning_hot_2019}. For older Sun-like stars, flaring events are less frequent but long-term variability cycles have been seen in the Sun, $\alpha$ Centauri AB, and Procyon, which change the X-ray--FUV flux by factors of 2--4 over several-year timescales \citep{ayres_2020_uv-xray}. It is therefore optimal to obtain simultaneous coverage of as much of the UV spectrum as possible.

Another important aspect is stellar multiplicity. While planet formation is more difficult in binary (or higher multiplicity) systems, it is not impossible \citep{kraus_2016_impact-of-multiplicity,moe_2021_binary-statistics}. To date, there are 759 confirmed planet-hosting multiple-star systems; this gives a multiplicity fraction for planet-hosting stars of $\sim22.5\%$ \citep{thebault_2025_binaries}. The HWO Preliminary Targets Catalog \citep{tuchow_2024_hpic} consists of $\sim13000$ targets which could be candidates in the search for habitable worlds---of these, roughly one third are in binary systems. The smaller list of 164 targets on the high-priority HWO ExEP Precursor Science Stars \citep{mamajek_2024_exep-list} also contains approximately 29\% binary systems\footnote{Determined simply by filtering the list for targets with an HIP component flag}. With this in mind, it is important to understand the impact that a stellar companion has on a planet's atmosphere. $\alpha$ Cen AB, consisting of a Sun-like G2 dwarf and sub-Solar K1 dwarf, provide an ideal laboratory to investigate the high-energy radiation environment in a Sun-like binary system. While neither star is confirmed to host a planet (although there have been proposed candidate planets, discussed at the end of \S \ref{sec:alpha-cen}), the close proximity and extensive observation history of the system allows for very accurate modeling to predict the environment around other more distant Sun-like binaries.

In this work, we present observations of the $\alpha$ Cen AB system from the Suborbital Imaging Spectrograph for Transition-region Irradiance from Nearby Exoplanet host stars (SISTINE) sounding rocket experiment, which provides FUV spectroscopy of exoplanet hosting stars and analogs. SISTINE covers a spectral bandpass of 980--1570 \ang, providing simultaneous measurements of several chromospheric and transition-region emission lines. We begin with a brief description of the $\alpha$ Cen system in section \ref{sec:alpha-cen}. In section \ref{sec:payload} we provide a description of the instrument design, assembly, and calibrations for the third flight of the SISTINE payload, designated as NASA mission 36.339 UG. Section \ref{sec:flight-results} discusses the flight results, including performance, data reduction, spectrum extraction, and stellar fluxes. Section \ref{sec:sed} discusses the creation of stellar spectral energy distributions (SEDs) for both stars. In section \ref{sec:planet} we investigate a hypothetical planet orbiting $\alpha$ Cen A, calculating the high-energy radiation incident on the planet and estimate atmospheric chemical abundances for the cases of a short-period gas giant and for an Earth-like terrestrial planet. Finally, section \ref{sec:summary} provides a brief summary of results from this work.

\section{The Alpha Centauri System}
\label{sec:alpha-cen}
The $\alpha$ Cen AB system consists of a G2 V star (A) and K2 V star (B). Table \ref{tab:stellar-properties} provides a summary of stellar properties and astrometry for the system. The two stars bracket the Sun in mass, radius, temperature, and luminosity. Despite being slightly larger and about 50\% more luminous, $\alpha$ Cen A is often considered a Solar-twin for several reasons: like the Sun, it is a slow rotator, with $P_{rot}\sim29$ d \citep{saar_osten_1997}; it has a nearly-Solar $\log{R'_{\rm{HK}}}$ of -5.00 \citep{henry_1996_caii}, indicating similar chromospheric activity levels; and its X-Ray activity levels are similar to that of the quiet Sun \citep{pagano_2004_stis,ayres_cycles_2023}.

The $\alpha$ Cen system is the nearest stellar system to the Sun---parallax measurements place it at a distance of 1.3319 pc ($\varpi=750.81$ mas) \citep{akeson_precision_2021}. The close proximity of the $\alpha$ Cen system is a boon for observations, allowing high signal-to-noise measurements in many wavelength bands. It also made direct EUV measurements possible with the Extreme Ultraviolet Explorer (EUVE), which made four observations of $\alpha$ Cen AB between 1993--1997 \citep{drake_1997_euve}. The close proximity, wealth of archival observations, and similarity of $\alpha$ Cen A to the Sun provide a unique opportunity to investigate the behavior of Sun-like stars and Sun-like binary systems.

\begin{deluxetable}{CcCc}[!ht]
\label{tab:stellar-properties}
\tablecaption{Properties of the $\alpha$ Centauri AB system}
\tablehead{\colhead{Parameter} & \colhead{Description} & \colhead{Value}    &\colhead{Reference}} 
\startdata
    a & Semi-major axis [arcsec] & 17.4930\pm0.0096 &   \citet{akeson_precision_2021} \\
    i   &   Inclination [deg]   &   79.2430\pm0.0089    &   \citet{akeson_precision_2021}\\
    P   &   Period [yr] &   79.762\pm0.019  &   \citet{akeson_precision_2021}\\
    e   &   Eccentricity    &   0.51947\pm0.00015   &   \citet{akeson_precision_2021}\\
    d   &   Distance [pc]   &   1.3319\pm0.0007 &   \citet{akeson_precision_2021}\\
    M_A   &   Mass of A [M$_\odot$]   &   1.0788\pm0.0029 &   \citet{akeson_precision_2021}\\
    M_B   &   Mass of B [M$_\odot$]   &   0.9092\pm0.0025 &   \citet{akeson_precision_2021}\\
    \mu   &   Mass fraction   &   0.54266\pm0.00011   &   \citet{akeson_precision_2021}\\
    R_A   &   Radius of A [R$_\odot$] &   1.2175\pm0.0055 &   \citet{akeson_precision_2021}\\
    R_B   &   Radius of B [R$_\odot$] &   0.8591\pm0.0036 &   \citet{akeson_precision_2021}\\
    L_A   &   Bolometric luminosity of A [L$_\odot$]  &   1.52 &   \citet{ayres_2020_uv-xray}\\
    L_B   &   Bolometric luminsosity of B [L$_\odot$] &   0.50 &   \citet{ayres_2020_uv-xray}\\
    T_A   &   Effective temperature of A [K]  &   5800\pm20   &   \citet{ayres_2020_uv-xray}\\
    T_B   &   Effective temperature B [K] &   5230\pm20   &   \citet{ayres_2020_uv-xray}\\
    $[Fe/H]$  &   Iron abundance  &   0.23    &   \citet{morel_chemical_2018}\\
    {P_{rot}}_A & Rotation period of A [days] &   29  &  \citet{saar_osten_1997}\\
    {P_{rot}}_B   &   Rotation period of B [days] &   42  &  \citet{saar_osten_1997}\\
    \log{R'_{\rm{HK}}}_A  &  $\log{R'_{\rm{HK}}}$ of A   &    -5.00   &   \citet{henry_1996_caii}\\
    \log{R'_{\rm{HK}}}_B  &  $\log{R'_{\rm{HK}}}$ of B   &   -4.92   &   \citet{henry_1996_caii}\\
\enddata
\end{deluxetable}

Both $\alpha$ Cen A and B have had claims of exoplanet detections \citep{dumusque_2012_alfcenb,wagner_2021_alfcena,beichman_2025_alfcena,sanghi_2025_alfcena}. The detections around $\alpha$ Cen A have not been confidently confirmed and remain candidate planets only, while the planetary signal detected around $\alpha$ Cen B is most likely a false detection \citep{hatzes_2013_alfcenb,rajpaul_2016_alfcenb}. However, \citet{zhao_2018_detectability} performed a noise analysis on an extensive archive of RV observations of the $\alpha$ Cen AB system from the Cerro Tololo Interamerican Observatory Echelle Spectrograph (CTIO) Echelle spectrograph, CTIO high resolution spectrograph, the High Accuracy Radial Velocity Planet Searcher (HARPS), and the Ultraviolet and Visual Echelle Spectrograph (UVES). Their study showed that, with the instruments considered, planets with a mass $M_p \lesssim53$ $M_\oplus$ would not be detectable in the habitable zone of $\alpha$ Cen A and planets with $M_p\lesssim8.4$ $M_\oplus$ would not be detectable in the habitable zone of $\alpha$ Cen B---thus, we cannot confidently rule out the existence of terrestrial or sub-Neptune type planets around either star.

\section{The SISTINE payload}
\label{sec:payload}
SISTINE provides FUV spectroscopy over a broad spectral bandpass with an in-flight resolving power of $R=\lambda / \Delta \lambda \sim 1500$ and spatial resolution of $\sim3''$. SISTINE also serves as a test bed for technological advances in FUV optical coatings and large-format detectors in support of future space-based UV observatories. The instrument comprises a $f/14$ Cassegrain telescope and a $2.1\times$ magnifying spectrograph. The spectrum is captured on a large-format microchannel plate (MCP) detector consisting of two $113\times42$ mm segments which span bandpasses of 980--1270 \ang\ on one segment and 1300--1580 \ang\ on the other. The wavelength range covers several stellar emission features with formation temperatures between 10$^{4-5.5}$ K which serve as chromospheric and transition-region diagnostics. A detailed description of the design, assembly and characterization of SISTINE is provided in \citet{nell_2024_sistine}.

\subsection{Assembly and Calibrations}
\label{subsec:calibrations}
Here we provide details of the calibrations performed prior to the third flight of SISTINE (SISTINE-3). Prior to flight we performed calibration tests at the vacuum UV facilities at the University of Colorado, Boulder. With the telescope and spectrograph fully assembled, we used a hollow cathode lamp to ionize ambient air and obtained a 2D spectrum which includes many bright emission features of nitrogen, hydrogen, and oxygen. This spectrum was cross-correlated with laboratory-measured wavelengths of the known emission lines and fit with a fourth-order polynomial to obtain the initial wavelength solution and characterize the instrument point spread function (PSF). Our pre-flight measurements showed a spectral resolving power of $R\sim1600$---consistent with the instrument resolving power being limited by the telescope PSF \citep{nell_2024_sistine}. Integration of the payload with the launch vehicle was performed at the NASA Wallops Flight Facility. More detailed descriptions of testing and calibrations prior to SISTINE-3 can be found in \citet{behr_sistine-3_2023} and details of the previous flight can be found in \citet{aguirre_radiation_2023}.

\section{SISTINE-3 flight results}
\label{sec:flight-results}
\begin{figure}[!ht]
    \centering
    \includegraphics[width=\linewidth]{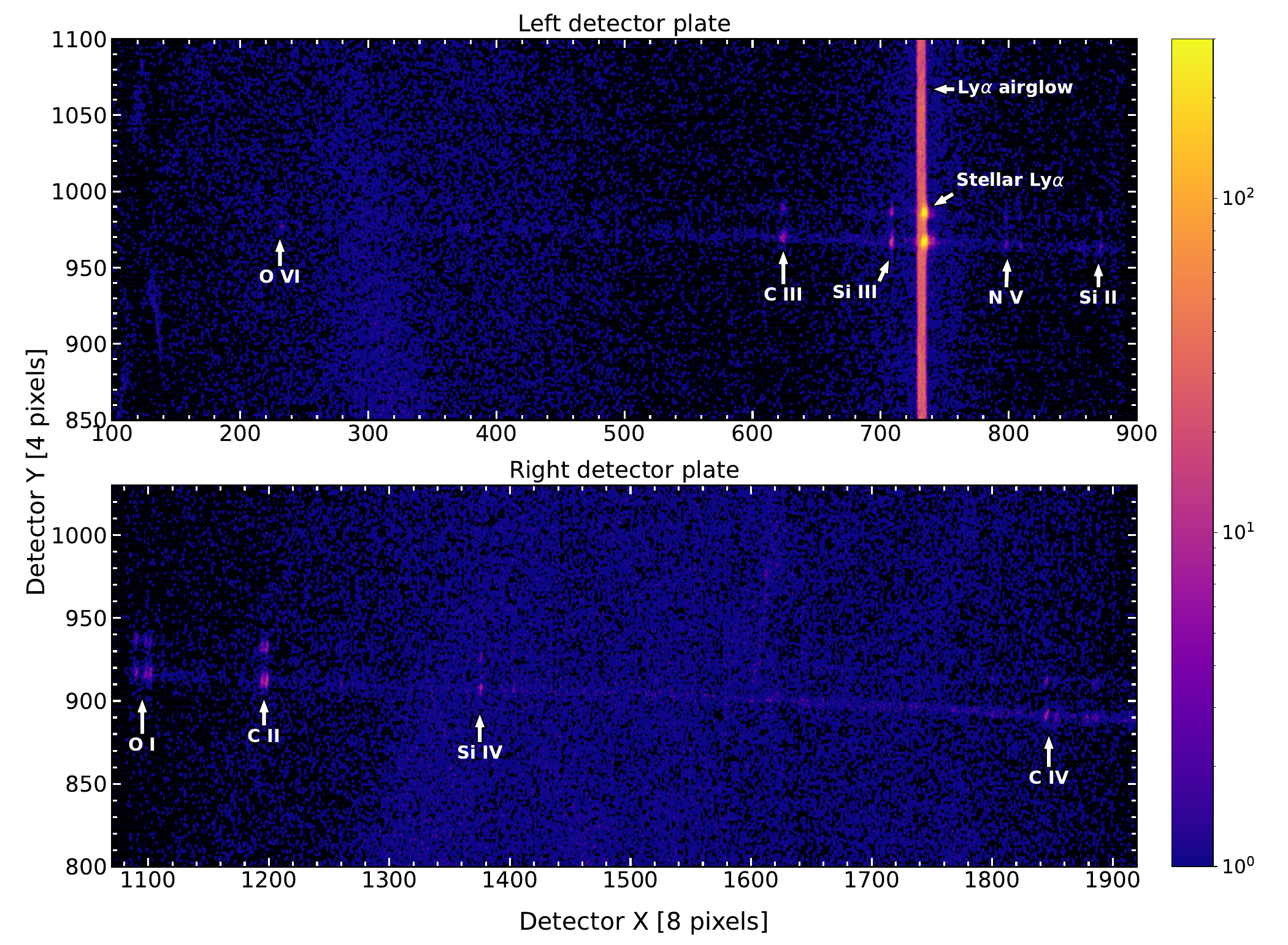}
    \caption{In-flight 2D spectrum of $\alpha$ Cen AB. The top panel shows the shorter wavelength detector segment (980--1270 \ang) and the bottom plot shows the longer wavelength segment (1300--1580 \ang). Data are shown in detector pixel coordinates and binned to 8 pixels in the X (dispersion) axis and 4 pixels in the Y (cross-dispersion) axis. For both panels the horizontal lines near the center of the image are the spectral traces of the two stars. The brighter bottom trace is $\alpha$ Cen A and the top trace is $\alpha$ Cen B. The bright vertical line at Lyman-$\alpha$ is geocoronal airglow which fills the instrument slit.}
    \label{fig:2d-flight}
\end{figure}

SISTINE-3 launched from Arnhem Space Center in the Northern Territory, Australia, on July 6$^{\text{th}}$, 2022 and successfully observed the $\alpha$ Cen AB system. Figure \ref{fig:2d-flight} shows the in-flight 2D spectrum from SISTINE-3 with annotations for some bright emission features. The figure is split into ``left plate'' and ``right plate'' which represent the two segments of the MCP detector. The bright vertical feature at Lyman-$\alpha$ is geocoronal airglow which fills the entire instrument slit. The background is highly variable in the dispersion and cross-dispersion directions; the excess emission is likely related to ionic field emission inside the instrument housing due to a relatively high payload pressure at the time of launch.

\begin{figure}[!ht]
    \centering
    \includegraphics[width=\linewidth]{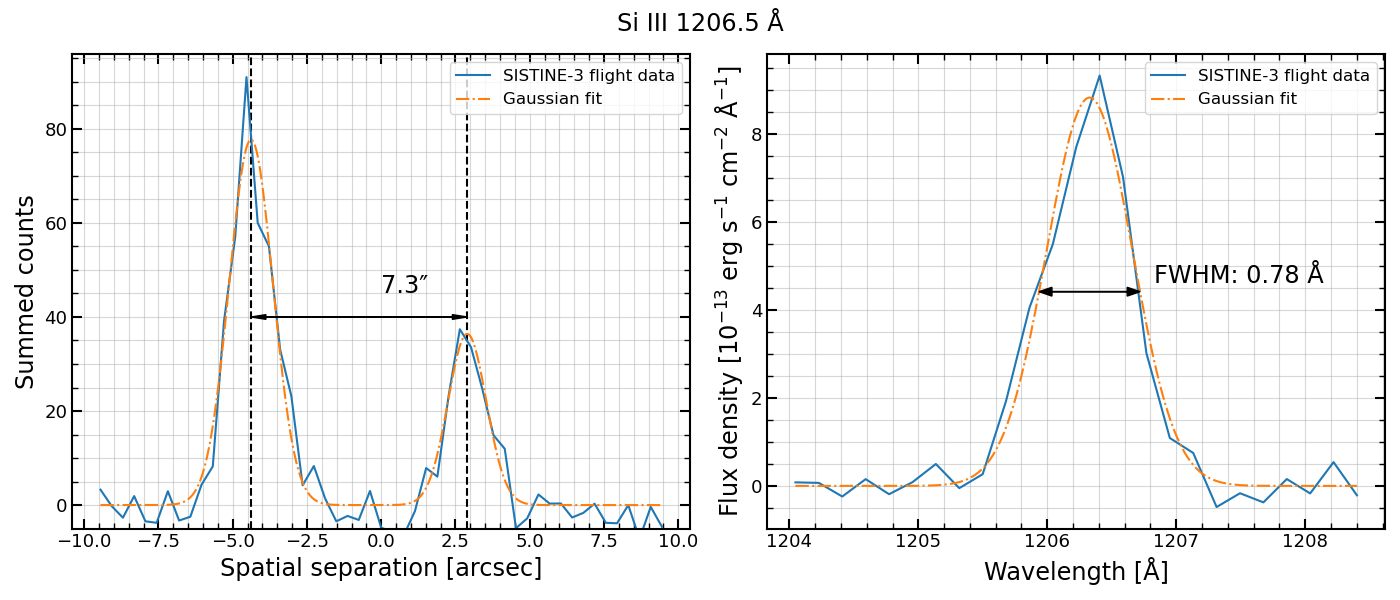}
    \caption{Spatial and spectral resolution of SISTINE-3 calculated using the Si III line at 1206.5 \AA. Left: Si III lines of $\alpha$ Cen A and B. The larger emission line is from A and the smaller from B. The lines are separated by $\sim7''$ which matches the predicted on-sky separation of the stars. Right: Si III emission of $\alpha$ Cen A.}
    \label{fig:resolution}
\end{figure}

The spectral trace of each star are seen as the horizontal lines crossing through the center of the image. The stars were spatially resolved and had a separation of $\sim7''$ on sky. Figure \ref{fig:resolution} shows an example of the spatial and spectral resolution in flight using the bright \ion{Si}{3} line at 1206.5 \AA. The in-flight spectral resolution was $R\sim1500$, consistent with pre-flight lab results.

\subsection{Instrument Drift}
\label{subsec:drift}
\begin{figure}[!ht]
    \centering
    \includegraphics[width=\linewidth]{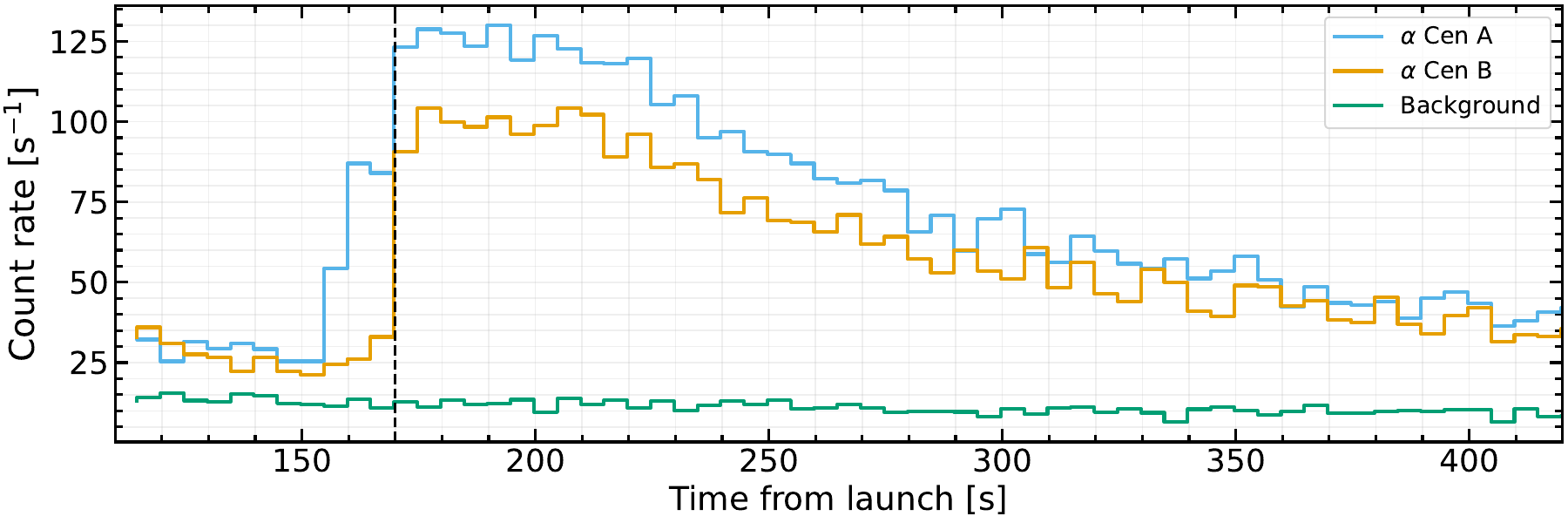}
    \caption{In-flight light curves from SISTINE-3. Data have been binned to 5 s intervals. The vertical black line shows when the payload reached its final pointing location. The count rate decreases over the duration of the flight due to the target drifting within the slit during observation.}
    \label{fig:lightcurve}
\end{figure}

Figure \ref{fig:lightcurve} shows the in-flight stellar count rate; the MCP detectors were turned on at 105 seconds and the payload came to its final on-target pointing position at $\sim$170 seconds. We obtained 250 seconds of on-target observation time. Observation of the stellar light curve shows a steady drop in count rate over the duration of observation, ending with a count rate of $\sim25\%$ that of the initial rate; this indicates that the targets drifted within the slit during the observation. The drift is likely due to thermal expansion from the payload being heated during launch. To quantify the amount of light lost we assumed a constant count rate equal to the average count rate over the initial time interval of 170-190 seconds. Taking the ratio of the total number of observed counts to the number predicted by the constant count rate we estimate that $\sim$42\% of the light from the stars was lost. Because the loss of light is assumed to be purely geometric in nature, we applied a constant correction factor to the entire spectrum.

\subsection{Spectrum Extraction}
\label{subsec:spectrum}
\begin{figure}[!ht]
    \centering
    \includegraphics[width=\linewidth]{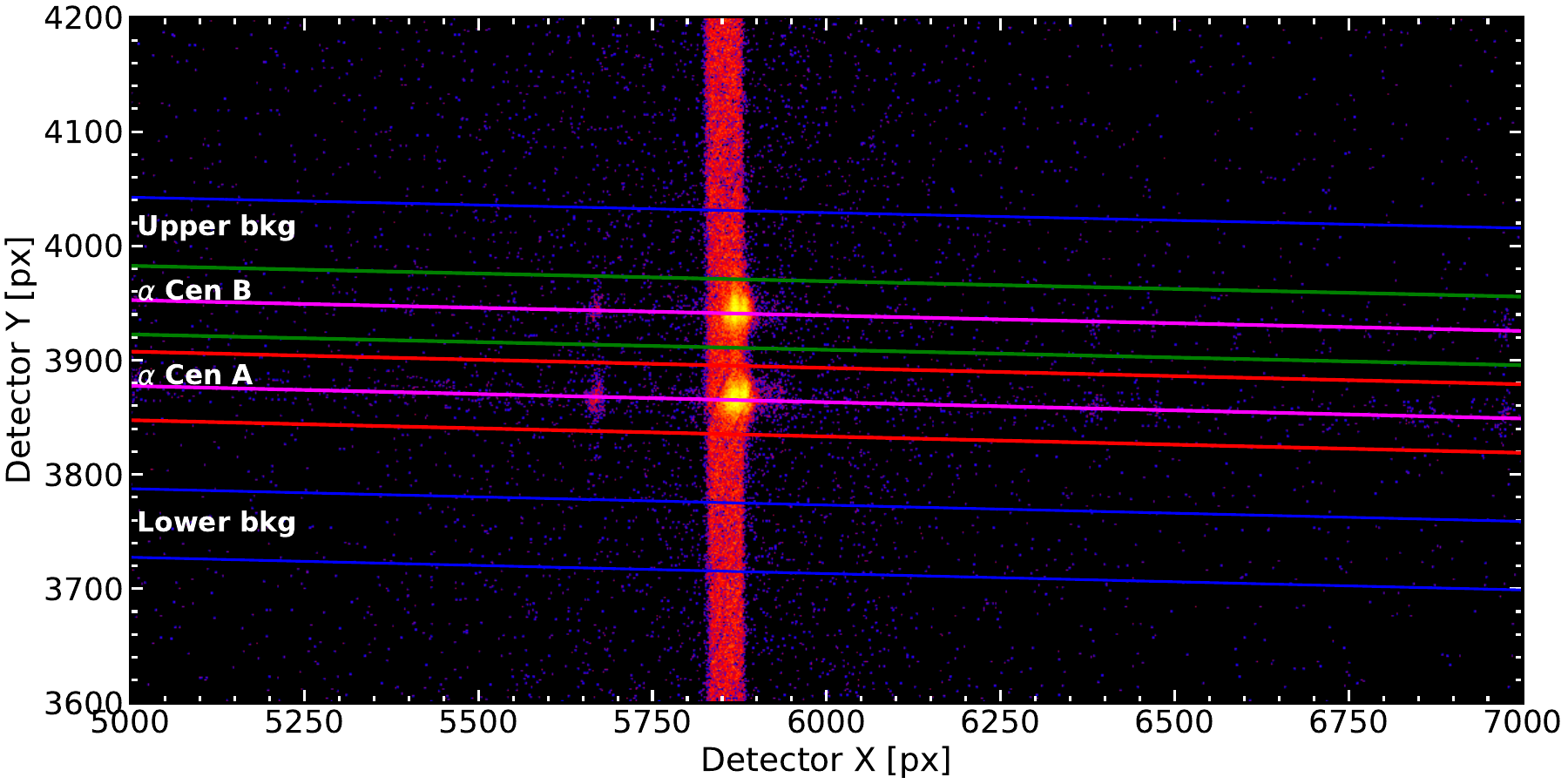}
    \caption{Example of the extraction process used to extract the stellar spectra of $\alpha$ Cen A and B from the full 2D flight image. The red lines bound the spectral region of $\alpha$ Cen A, and the green lines the spectral region of $\alpha$ Cen B. The dark blue lines define the background regions. Magenta lines show the linear fit to the spectral trace.}
    \label{fig:extraction}
\end{figure}

\begin{figure}[!ht]
    \centering
    \includegraphics[width=\linewidth]{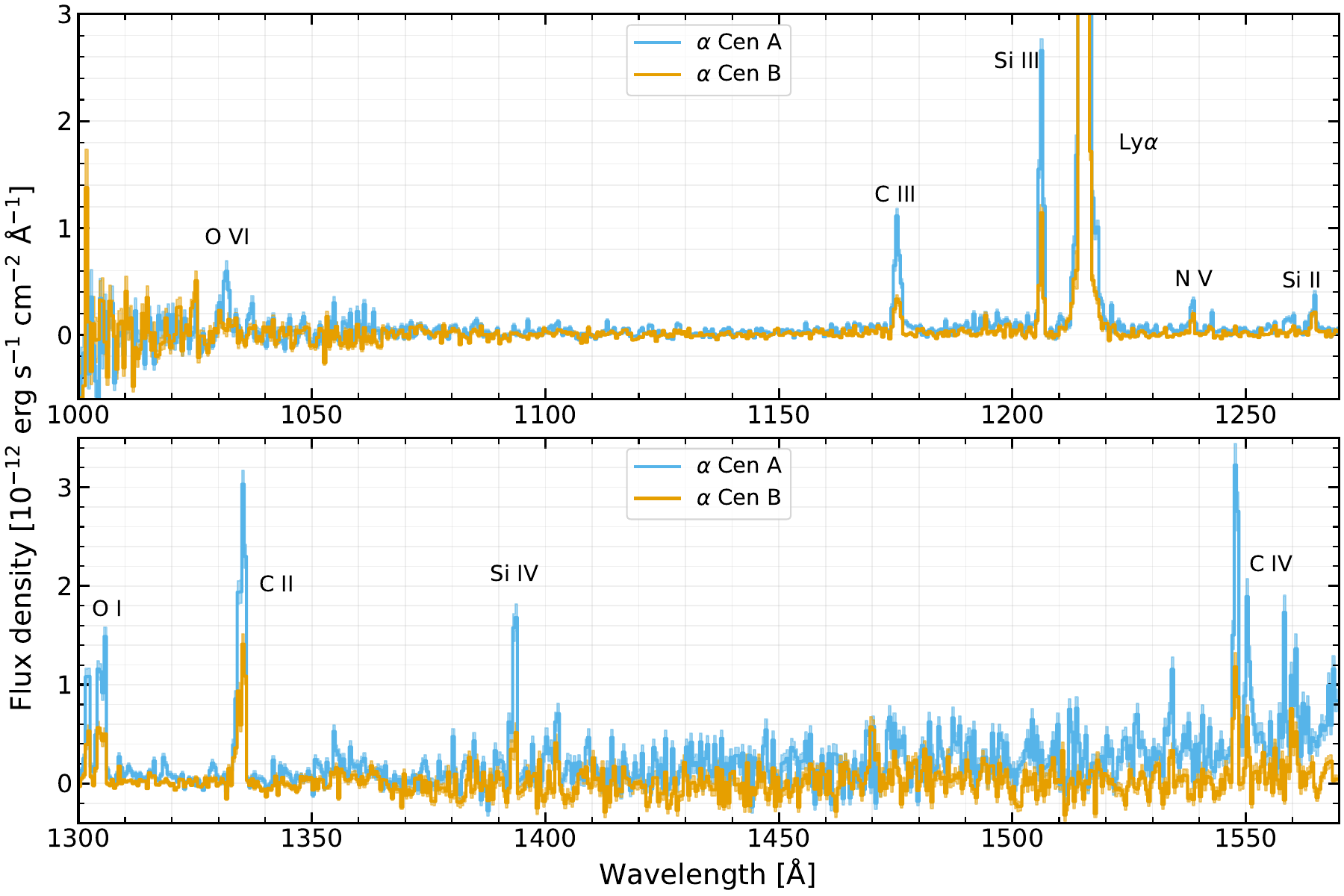}
    \caption{FUV spectrum of $\alpha$ Cen A (blue) \& B (orange) observed by SISTINE-3. 1-$\sigma$ errors are shown as a filled areas with light transparency. Data are binned to 0.5 \ang.}
    \label{fig:spec-1d}
\end{figure}

After applying the correction for loss of light, we extracted the stellar spectra using the following procedure. We fit a line to the spectral trace of each star using the centroids of two bright emission features as reference points; this is done separately on each detector segment because the spectra are offset between segments. Using the linear fits to the spectral trace, we made a region extending 30 pixels above and below the spectral trace and took the sum in the cross-dispersion direction to create a 1D stellar spectrum. We performed a background subtraction by making two 60 pixel wide regions above and below the the spectral traces and summed the counts in both regions. We then scaled the total background counts to match the width of the spectral extraction region and subtracted it from the 1D spectrum on a pixel-by-pixel basis. Figure \ref{fig:extraction} shows the spectral and background extraction regions for one detector segment overlaid on the 2D flight image. There is some excess emission below the \ion{Si}{3} and \lya\ features for $\alpha$ Cen A due to the spectral trace moving on the detector because of the drift (\S \ref{subsec:drift}). The lower background region was shifted downward to avoid including the excess emission during background subtraction. Figure \ref{fig:spec-1d} shows the background-subtracted 1D spectrum. The reflectance of the optics coatings drops rapidly for wavelengths less than 1050 \ang\ \citep[see Figure 4 of][]{nell_2024_sistine} leading to poor signal-to-noise in that region.

\subsection{Stellar Emission Lines}
\label{subsec:emission-lines}
We measured fluxes for 7 emission lines across the entire SISTINE-3 bandpass; \ion{O}{6} ($\lambda\lambda$1032,1038 Å), \ion{C}{3} ($\lambda$1175 Å), \ion{Si}{3} ($\lambda$1206 Å), Ly$\alpha$ ($\lambda$1216 Å), \ion{N}{5} ($\lambda$$\lambda$1238,1243 Å), \ion{Si}{4} ($\lambda$$\lambda$1394,1403 Å), and \ion{C}{4} ($\lambda$$\lambda$1548,1551 Å). A list of all the emission line flux values for $\alpha$ Cen A and B is provided in Table \ref{tab:fluxes}. The O VI emission from $\alpha$ Cen B was below the noise level of our observations and we report the root mean square of the flux over the emission line region as an upper limit.

\begin{deluxetable}{ccccc}[!ht]
    \label{tab:fluxes}
    \tablecaption{Emission line fluxes for $\alpha$ Cen AB and the Sun}
    \tablehead{\colhead{Feature} & \colhead{Wavelength} & \colhead{$\alpha$ Cen A Flux}    &\colhead{$\alpha$ Cen B Flux}   &   \colhead{Scaled Solar Flux}\\
        &   \colhead{(\AA)}   &   \colhead{($10^{-13}$ erg s$^{-1}$ cm$^{-2}$ \AA$^{-1}$)}  &   \colhead{($10^{-13}$ erg s$^{-1}$ cm$^{-2}$ \AA$^{-1}$)}  &   \colhead{($10^{-13}$ erg s$^{-1}$ cm$^{-2}$ \AA$^{-1}$)}} 
    \startdata
        \ion{O}{6}  &   1032,1037   &   $3.82\pm0.72$ &   $<0.92$   & ---\\
        \ion{C}{3} & 1175   &   $8.39\pm0.29$  &   $5.33\pm0.32$ &   $7.45\pm0.12$\\
        \ion{Si}{3}  &   1206    &   $15.6\pm0.5$ &   $9.34\pm0.46$  &   $13.7\pm0.6$\\
        \ion{H}{1} Ly$\alpha$  &  1216 &   $864\pm27$  &   $1289\pm87$ &   $904\pm3$\\
        \ion{N}{5} &  1238,1243    &   $3.70\pm0.21$  &   $2.76\pm0.26$ &   $2.63\pm0.04$\\
        \ion{C}{2}  &   1335    &    $26.9\pm0.7$   &   $21.3\pm0.8$   &   $23.9\pm0.6$\\
        \ion{Si}{4}   &   1394,1403   &   $11.3\pm0.5$    &   $8.01\pm0.90$ &   $9.26\pm0.20$\\
        \ion{C}{4}    &    1548,1551  &   $22.6\pm0.8$    &   $19.6\pm1.50$ &   $18.8\pm0.3$\\
    \enddata
    \tablecomments{\ion{H}{1} Ly$\alpha$ for SISTINE is reported as the integrated flux of the reconstructed line from 1214--1217 \AA. All solar fluxes are reported as the average of the min and max fluxes from SOLSTICE tabulated in \citet{ayres_2020_uv-xray} and scaled to the distance of the $\alpha$ Cen AB system.}
\end{deluxetable}

\section{Full SED of Alpha Centauri A and B}
\label{sec:sed}
\begin{figure}[!ht]
    \centering
    \includegraphics[width=0.93\linewidth]{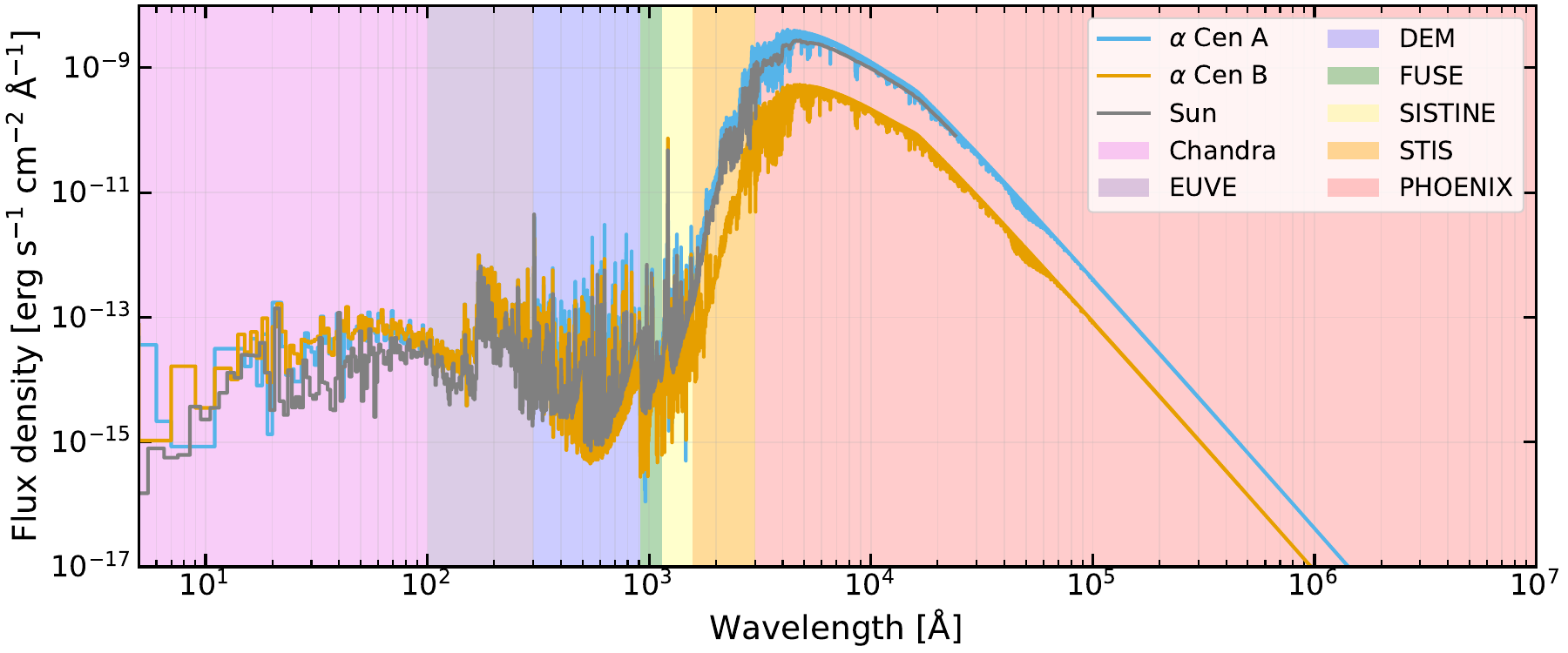}
    \caption{Stellar SEDs for $\alpha$ Cen A (blue) and B (orange) binned to 1 \ang. The SED of the Sun, scaled to a distance of 1.33 pc, from \citet{woods_solar_2009} is shown in gray for comparison. Colored background regions show the various regions described in \S \ref{subsec:SED}. The regions containing the reconstructed Ly$\alpha$ (1214--1217 \AA, \S \ref{subsubsec:lya}) and SISTINE detector gap (1270--1300 \AA, \S \ref{subsubsec:fuv-nuv}) are not colored for clarity. The \texttt{PHOENIX} model extends to $10^7$ \AA\ but we have restricted the Y axis to $10^{-17}$ erg s$^{-1}$ cm$^{-2}$ $\mathrm{\AA}^{-1}$, again to aid visual clarity of features within the main spectra.}
    \label{fig:full-sed}
\end{figure}

We constructed a full spectral energy distribution (SED) for both $\alpha$ Cen A and B using the SISTINE-3 FUV spectrum combined with archival observations from several other observatories in the X-ray,  EUV, FUV, and NUV, and a stellar atmosphere model for the visible-IR. Descriptions of the observations or models used for each spectral region are provided in \S \ref{subsec:SED}. The SED spans 5 \ang--1 mm in wavelength. Figure \ref{fig:full-sed} shows the full SED of both stars. Due to the luminosity variability of $\alpha$ Cen A and B and because our data were not taken contemporaneously, we created a variability model (\S \ref{subsec:variability}) to scale flux from observations such that each SED component is adjusted to reflect the star's variability state during the SISTINE-3 observations, which was coincidentally near minimum for both stars in 2022. In this section we first describe the variability model and then provide a brief description of the data used in each section of the final SED.

\subsection{Variability Model}
\label{subsec:variability}
Decades of observations of the $\alpha$ Cen AB system have shown that the X-ray and FUV luminosity of both stars varies cyclically, similar to the Sun, with periods of $\sim19$ yr for $\alpha$ Cen A and $\sim9$ yr for $\alpha$ Cen B \citep{ayres_2014_cycles,ayres_cycles_2023}. Because the various observations that we used to create the SEDs were taken at various times ranging from 1993--2022, there can be large differences in the observed flux due to the cyclic activity levels. We corrected for this by creating a variability model which is both time- and wavelength-dependent and used it to scale each observation to a common activity level which we chose to be the activity level in 2022 at the time of the SISTINE-3 observation.

\begin{figure}[h]
    \centering
    \includegraphics[width=0.93\linewidth]{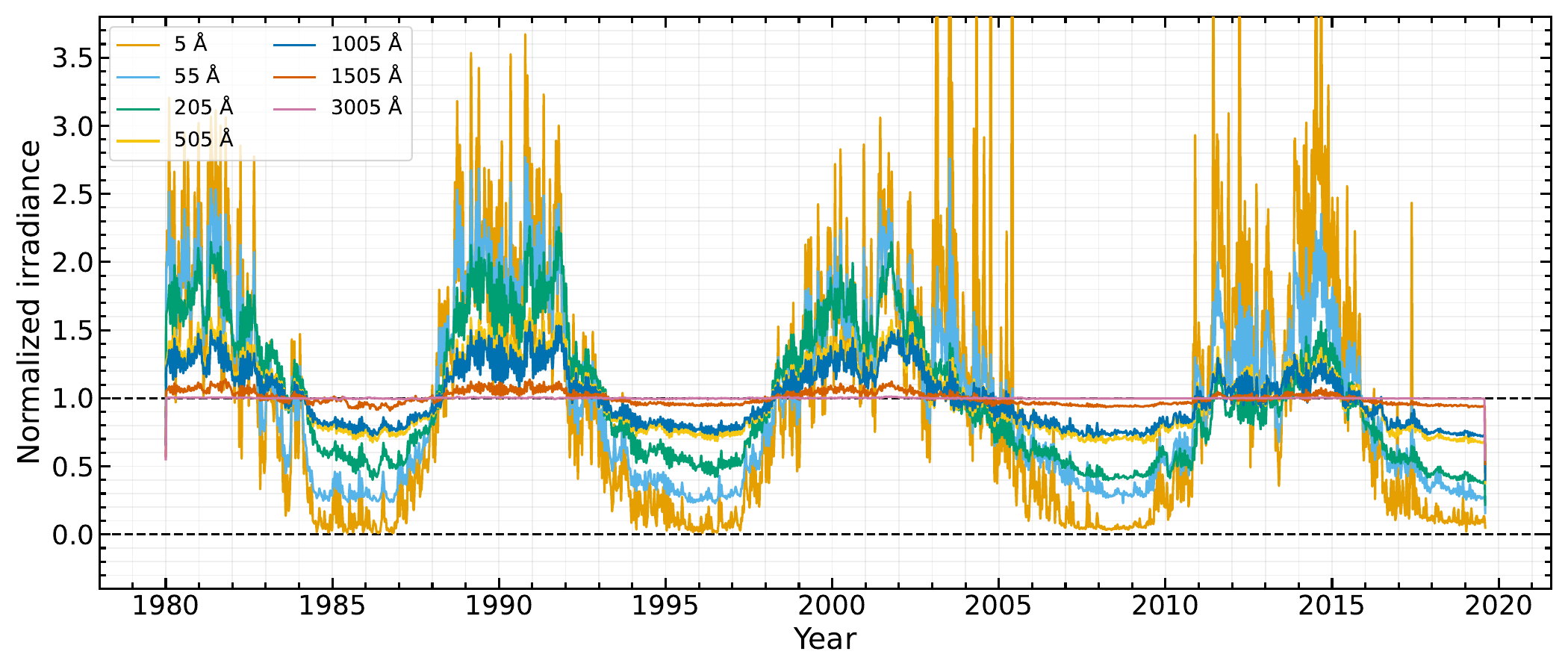}
    \caption{Solar irradiance normalized to the average value. Data are shown as a 10 day rolling average normalized to the average of the entire time series. Data were retrieved from the LISIRD database.}
    \label{fig:solar-time-series}
\end{figure}

Because of their similar X-ray and chromospheric activity level indicators (i.e., rotation period and $L_X/L_{bol}$ \citep{wright_2011_activity} and $\log{R'_{HK}}$ \citep{henry_1996_caii}), we make the assumption for both $\alpha$ Cen A and B that the wavelength-dependence of variability behaves similarly to the Sun. Using data archived on the LASP Interactive Solar Irradiance Datacenter\footnote{\url{https://lasp.colorado.edu/lisird/}} (LISIRD), we created a wavelength-dependent variability model for the Sun. The LISIRD database provides solar irradiance measurements at a daily cadence. Figure \ref{fig:solar-time-series} shows a 10 day rolling average of the Solar irradiance normalized to the average irradiance over the 40 year time span for several wavelengths.

\begin{figure}[!ht]
    \centering
    \includegraphics[width=\linewidth]{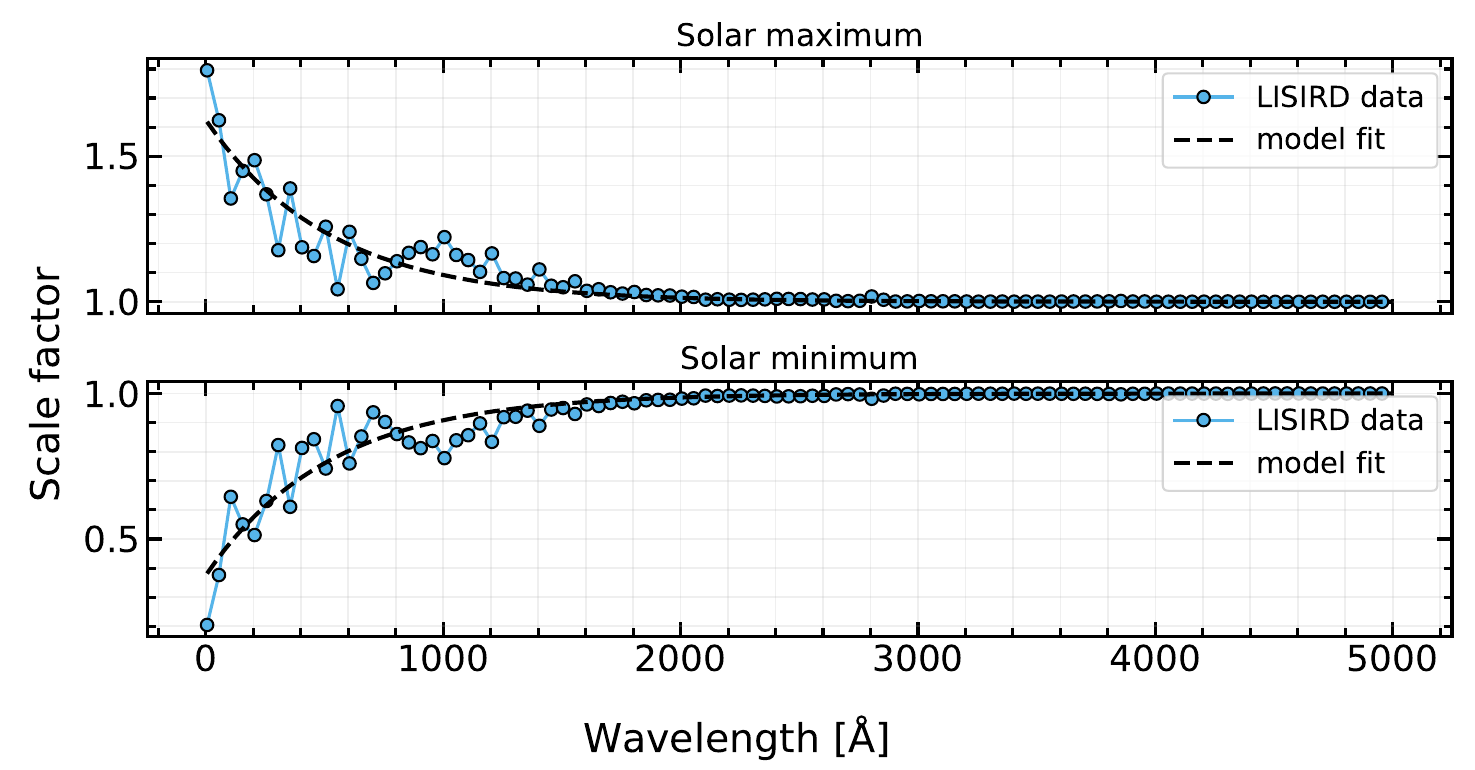}
    \caption{Functional fits to the Solar activity cycle as a function of wavelength. The scale factors are normalized such that a factor of 1 represents the Sun at an average activity level between Solar maximum and minimum.}
    \label{fig:variability}
\end{figure}

To create our variability model, we took the average irradiance measurements over a window of 3 years centered on each solar maximum or minimum. We then normalized each average maximum and minimum to the average irradiance over the entire 40 year time series in 100 \ang\ intervals. A plot of the resulting variability model is shown in Figure \ref{fig:variability}. We find that variability is largest at short wavelengths and decreases at longer wavelengths until the variability becomes insignificant longward of $\sim2200$ \ang.

We find a good fit to the data with an exponential of the form

\begin{equation}
    s=1\pm0.6185e^{-0.0191(\lambda-5)}
\end{equation}

where $s$ is the scale factor and $\lambda$ is the wavelength in nm.

\subsection{SED creation}
\label{subsec:SED}
The full SEDs of $\alpha$ Cen A and B were created using a combination of new and archival observations and synthetic models. We have used archival observation from the Chandra X-ray Observatory, Far-Ultraviolet Spectroscopic Explorer (FUSE), EUVE, and HST, the new observations reported in this paper from SISTINE, and model spectra using \texttt{PHOENIX} models and DEMs. Each observation or model is described in the following sections corresponding to the wavelength regime in which it was used.

\subsubsection{X-ray}
\label{subsubsec:x-ray}
\begin{figure}[!ht]
    \centering
    \includegraphics[width=\linewidth]{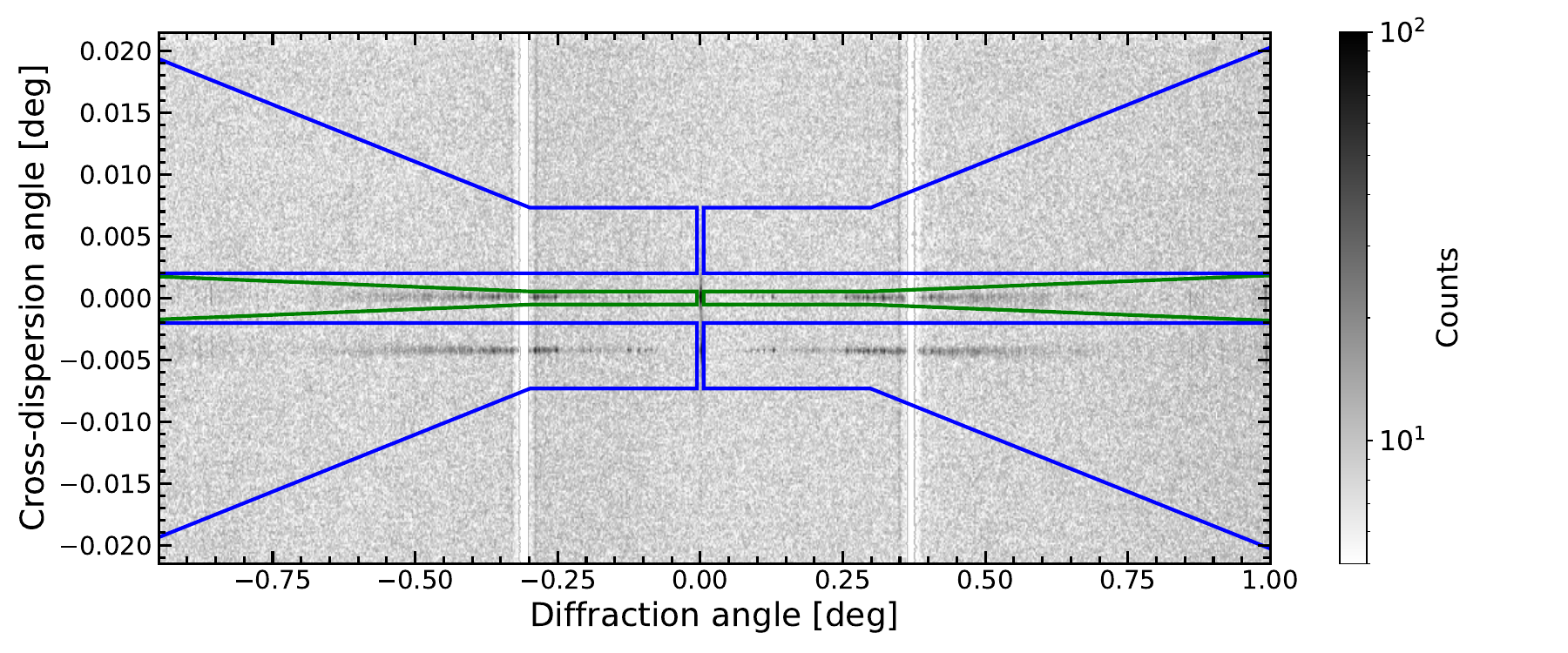}
    \caption{2D LETGS spectrum of $\alpha$ Centauri A \& B. The upper spectral trace is $\alpha$ Cen A and the lower is $\alpha$ Cen B. The green and blue regions show the spectrum extraction region and background extraction regions, respectively. The two empty vertical stripes are gaps between the detector plates which contain no data. Negative diffraction angle is order $m=-1$ and positive is $m=1$. Note that while the stars are well resolved, the lower background region for $\alpha$ Cen A entirely encloses $\alpha$ Cen B. Likewise, the upper background region for $\alpha$ Cen B (not shown) encloses the spectral trace of A. During extraction of the 1D spectra, we moved the background region of $\alpha$ Cen A (B) down (up) to ensure that none of the spectral trace of the other star remained within the background region.}
    \label{fig:letgs-default-extraction}
\end{figure}

We retrieved X-ray observations of both $\alpha$ Cen A and B from Chandra using the Low Energy Transmission Grating and High Resolution Camera spectroscopic array detector (LETGS). We found three archived observations of $\alpha$ Centuari in the LETGS configuration: obsid 29 (1999-12-24; PI Brinkman), 7432 (2007-06-04; PI Ayres), and 12332 (2011-06-08; PI Ayres). We used obsid 29 because it had the largest spatial separation of the two stars, allowing us to extract the two stellar spectra with no overlap. The standard Chandra pipeline was not able to successfully extract spectra of the stars because the background region of one star overlapped with the spectral region of its companion. Figure \ref{fig:letgs-default-extraction} shows an example of the default extraction regions for $\alpha$ Cen A. The lower background region entirely encloses the spectral trace of $\alpha$ Cen B. Likewise, for $\alpha$ Cen B, the upper background region encloses the spectral trace of $\alpha$ Cen A.

We re-extracted the spectra of each star using custom background regions which shift either the upper ($\alpha$ Cen B) or lower ($\alpha$ Cen A) background region up or down such that it no longer encloses the spectral trace of the other star. We then used the Chandra Interactive Analysis of Obseravtions (\texttt{CIAO v4.17}) software to extract a spectrum using the updated background regions. The X-ray spectra of both $\alpha$ Cen A and B are dominated by 

\subsubsection{EUV}
\label{subsubsec:euv}
The close proximity of the $\alpha$ Cen system made direct EUV observations possible with EUVE and the system was observed in 1993, 1995, and 1997 \citep{drake_1997_euve}. EUVE had a large aperture and modest angular resolution and therefore was unable to spatially resolve the individual stars; because of this, each EUVE observation contains flux from both $\alpha$ Cen A and B. Based on the results of \citet{drake_1997_euve} and \citet{duvvuri_2025_euv}, we assume an even contribution from $\alpha$ Cen A and B and use the same EUVE spectrum with a scaling factor of 0.5 for each star. We use the EUVE spectrum with an ISM attenuation correction \citep[see Section 3.6 of][]{aguirre_radiation_2023} for wavelengths between 100--300 \ang.

\begin{figure}[!ht]
    \centering
    \includegraphics[width=\linewidth]{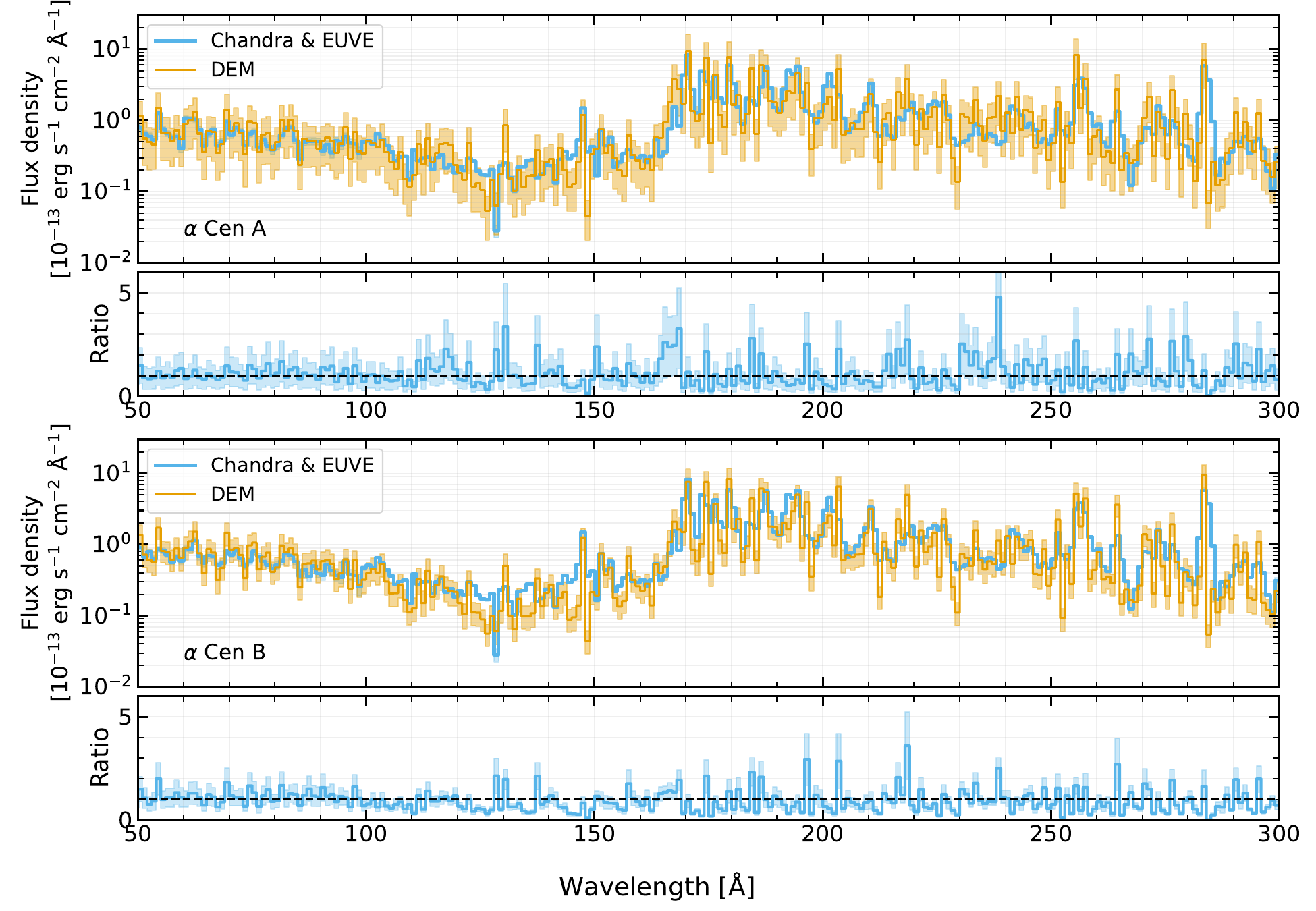}
    \caption{Comparison of the $\alpha$ Cen A (top) and B (bottom) DEM flux to the observed flux from Chandra and EUVE. Each two-panel subplot is laid out as follows: \textbf{Top:} Synthetic XUV spectrum from the DEM (orange) compared with the observed $\alpha$ Cen spectrum from Chandra LETGS and EUVE (blue). All data are binned to 1 \ang. \textbf{Bottom:} Ratio of DEM flux to observed flux.}
    \label{fig:dem-fit}
\end{figure}

The sensitivity of EUVE is low for wavelengths longer than 300 \AA, so we use a synthetic model spectrum created from a differential emission measure technique for wavelengths between 300--910 \AA\ \citep{duvvuri_2025_euv}. The DEM method assumes an optically thin, collisionally dominated plasma in which the flux of an emission line is given by the temperature integral of the product of a contribution function and the emission measure. Observed fluxes from optically thin FUV and X-ray coronal and transition-region lines (formation temperatures of $T \sim 10^4$-–$10^7$ K) are used to infer the plasma’s temperature–density structure and predict fluxes in unobserved lines \citep{mariska_1992_transition,warren_1998_dem,duvvuri_reconstructing_2021}. In constructing the $\alpha$ Cen DEMs, the likelihood function used in the fitting process was defined as the sum of three terms: two comparing DEM-predicted and observed fluxes for each star, and one comparing their combined predicted flux to the unresolved EUVE spectrum \citep{duvvuri_2025_euv}. Figure \ref{fig:dem-fit} shows the EUV spectrum produced by the DEM technique compared to the observed Chandra and EUVE spectra from 50--300 \AA. The ratio of integrated flux (DEM/observed) over this region is $1.01\pm0.07$ for $\alpha$ Cen A, $0.94\pm0.04$ for $\alpha$ Cen B, and $0.98\pm0.05$ for the combined spectrum. The X-ray and FUV emission lines included in the DEM fitting process are tabulated in \citet{duvvuri_2025_euv}.

\subsubsection{FUV \& NUV}
\label{subsubsec:fuv-nuv}
The FUV portion of the SED is covered by a combination of FUSE, SISTINE, and ASTRAL spectra\footnote{\url{https://casa.colorado.edu/~ayres/ASTRAL/}} \citep{ayres_2013_astral}. ASTRAL spectra consist of the coaddition of many HST-STIS echelle observations using the E140M and E140H gratings for the FUV and the E230M and E230H gratings for the NUV. Use of the SISTINE spectrum in the FUV is desirable because it provides a reference point for the scaling relations since the observation was taken while both stars were at their minimum activity level and because it allows us to use a single scaling relation for the entire FUV region rather than scaling the many individual observations from HST that are required to cover the full FUV bandpass.

For wavelengths between 910--1140 \ang\ we use FUSE data. We retrieved two observations of both $\alpha$ Cen A and B taken on May 5--6, 2006, using the FUSE medium resolution aperture ($4''\times20''$) which is capable of spatially resolving the pair. The spectra were weighted by exposure time and coadded to create the final product used for the SEDs. For wavelengths between 1140--1565 \ang\, we use the flux-calibrated spectrum from SISTINE---except in the \lya\ emission line, which is discussed in \S \ref{subsubsec:lya} and the gap between the SISTINE detector plates (1270--1300 \ang). While we were able to obtain enough signal to measure the flux in the \ion{O}{6} ($\lambda\lambda1032,1037$) line for $\alpha$ Cen A, the signal in the continuum of the SISTINE spectra drop rapidly after $\sim1070$ \ang\ for both stars. For this reason we opted to use the FUSE spectrum until 1140 \ang\ to provide good signal-to-noise within the continuum region, and the SISTINE spectrum from 1140--1565 \ang\ to provide simultaneous coverage of the major emission features within this region. For the remainder of the FUV and NUV wavelengths from 1565--3000 \ang, and the gap between SISTINE detector plates, we us spectra from the ASTRAL database. Because the ASTRAL spectra consist of many coadded STIS observations, the observation times span a broad range from 1999--2017, which is comparable to the activity cycle period of $\alpha$ Cen A and nearly twice that of B. We opted not to scale individual exposures of the ASTRAL spectra for two reasons: first, ASTRAL spectra already contain scale factors used to scale individual exposures to that with the highest throughput. The $\alpha$ Cen A spectra are scaled to match the observation taken 1999-02-12 and the $\alpha$ Cen B spectra to the observation taken on 2010-07-01. Second, the ASTRAL spectra are used only for wavelengths longer than 1565 \ang, after which we predict the flux variability to be less than 3\%, which is comparable to or smaller than the uncertainty on the flux observations themselves.

\subsubsection{Lyman-$\alpha$}
\label{subsubsec:lya}
\begin{figure}[!ht]
    \centering
    \includegraphics[width=\linewidth]{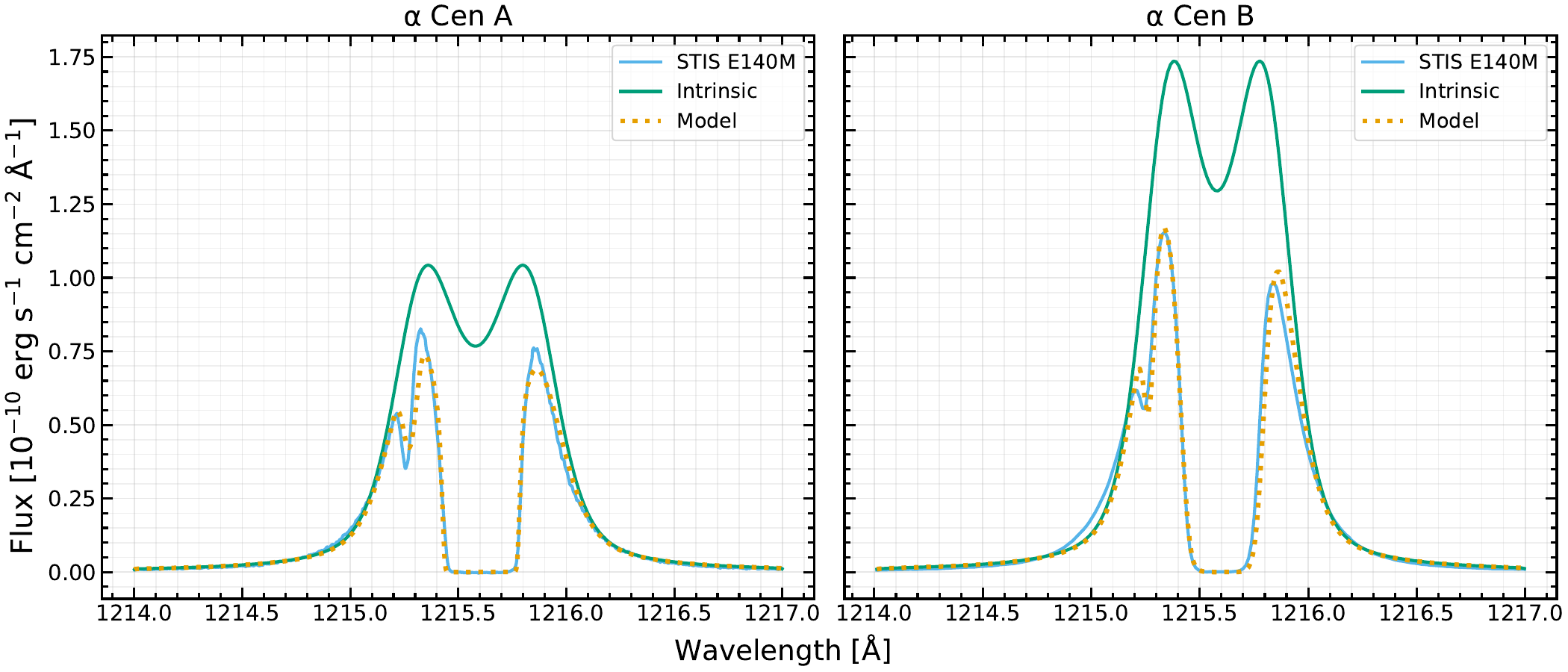}
    \caption{Reconstructed \lya\ emission line profiles for $\alpha$ Cen A (left) and B (right). The blue line represents the observed data from HST-STIS E140M, the green line the reconstructed intrinsic profile from the \citet{youngblood_intrinsic_2022} model, and the dashed orange line the model fit to the observed data.}
    \label{fig:lya}
\end{figure}

\lya\ emission is heavily attenuated by neutral hydrogen in the ISM; often, the core of the line is completely saturated even for the nearest stars, leaving only the wings of the emission line observable. We follow the approach of \citet{youngblood_intrinsic_2022} to reconstruct the intrinsic \lya\ profile from existing HST spectra. The \citet{youngblood_intrinsic_2022} model simultaneously fits a self-absorbed Voigt profile and ISM absorption to the HST data. Free parameters include intrinsic flux, \ion{H}{1} and \ion{D}{1} column densities, ISM Doppler parameter, stellar and ISM radial velocities (RV), and a self-reversal parameter. We fix the values of the systemic RV and ISM parameters to those reported in \citet{linsky_1996_ism} and also add an additional fixed absorption component to account for extra absorption from the heliosphere. Figure \ref{fig:lya} shows the reconstructed \lya\ profile for both $\alpha$ Cen A and B, which is used for wavelengths between 1214--1217 \ang.

\subsubsection{Visible \& Infrared}
\label{subsubsec:ir}
Finally, for visible--IR wavelengths between 3000--10$^7$ \ang, we use a \texttt{PHOENIX} model from the BT-Settl CIFIST grid \citep{baraffe_2015_model,allard_2016_phoenix} retrieved from the SVO Theoretical Spectra web database\footnote{\url{https://svo2.cab.inta-csic.es/theory/newov2/index.php?models=bt-settl-cifist}}. Model spectra are provided on a grid of $T_\mathrm{eff}$ and $\log{g}$ which have steps of 100 K and 0.5 cm s$^{-2}$, respectively. 

We selected the model spectra closest to the stellar parameters listed in Table \ref{tab:stellar-properties} and interpolated them onto a common grid following the method described by \citet{wilson_2021_trappist}. For $\alpha$ Cen A, we employed two bounding models with $(T_{\mathrm{eff}}, \log g) = (5800, 4.0)$ and $(5800, 4.5)$. For $\alpha$ Cen B, we used four models with ($T_{\mathrm{eff}},\log{g})$ = (5200, 4.0), (5200, 4.5), (5300, 4.0), and (5300, 4.5).

Flux from the \texttt{PHOENIX} models is measured at the stellar surface so we first scale by the squared ratio of stellar radius and distance, then we apply a second scaling factor to accurately match the distance-scaled \texttt{PHOENIX} model to the HST spectra between 2000--3000 \ang\ using a least squares method.

\section{Simulating an Earth-like planet in a Sun-like binary system}
\label{sec:planet}
Orbital dynamics simulations of the $\alpha$ Cen AB system have shown that there are stable orbits in either a circumbinary ``p-type'' orbit around both stars or a circumstellar ``s-type'' orbit around either of the individual stars \citep{wiegert_1997_alpha-cen,rabl-dvorak_1998_binaries,holman_1999_binary,quarles_2020_binary}. The stability of an orbit is defined by whether the planet is inside the ``critical semi-major axis''---the largest semi-major axis at which all test particles in the simulation survived the full integration time. Using the expression for critical semi-major axis defined in \citet{holman_1999_binary}, a planet orbiting $\alpha$ Cen A in a prograde, coplanar, circular orbit with zero inclination is stable out to $\sim$2.5 AU on Myr time scales. The more recent simulations by \citet{quarles_2020_binary} place the critical semi-major axis for $\alpha$ Cen A just slightly farther at 2.78 AU and also show that if the orbit is retrograde, the critical semi-major axis extends to 3.84 AU.

In this section, we use the SEDs created for $\alpha$ Cen A and B as inputs to the \texttt{VULCAN} photochemical kinetics code \citep{tsai_2021_vulcan} to model the atmosphere of a hypothetical Earth-like planet orbiting $\alpha$ Cen A. \citet{kaltenegger_calculating_2013} calculate the habitable zone for $\alpha$ Cen A to be between 0.924--2.194 AU. We set our hypothetical planet on a circular orbit at a distance of 1.2 AU from $\alpha$ Cen A---the distance at which the planet would, on average, receive similar instellation as Earth receives from the Sun, $\sim1361$ W m$^{-2}$.

\subsection{Incident XUV flux}
\label{sec:xuv-flux-sim}

\begin{figure}[!ht]
    \centering
    \includegraphics[width=0.8\linewidth]{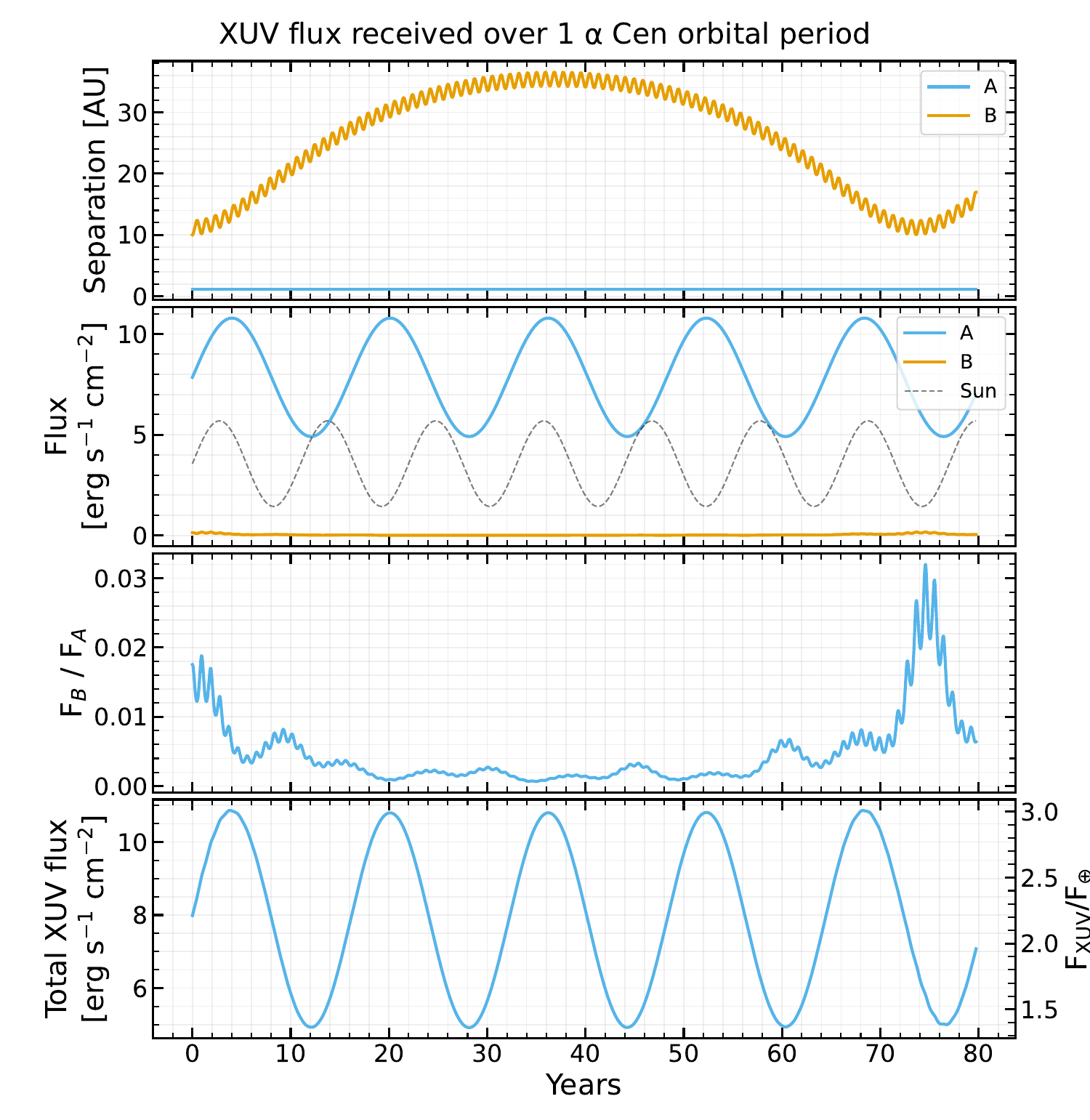}
    \caption{XUV ($\lambda<912$ \ang) flux received by a hypothetical planet orbiting $\alpha$ Cen A at 1.2 AU over the course of 1 stellar orbital period. The panels show, from top to bottom: \textbf{First:} Solid lines show the distance of the planet from $\alpha$ Cen A and B. The separation from $\alpha$ Cen A is constant at 1.2 AU. \textbf{Second:} The XUV flux received at the planet from each star. The dashed gray line shows the XUV flux that Earth receives from the Sun for comparison. \textbf{Third:} The ratio of flux received from $\alpha$ Cen B compared to $\alpha$ Cen A. \textbf{Fourth:} The left side axis shows total XUV flux received from both stars. The right side axis shows the ratio of the total XUV flux at the planet compared to the average XUV flux Earth receives from the Sun.}
    \label{fig:orbit-sim}
\end{figure}

As X-ray and EUV (XUV; $\lambda<912$ \ang) flux is the dominant driver of upper atmospheric heating and escape \citep{lammer_2003_escape,tian_2015_atmospheric_escape,chadney_2015_mass-loss}, we begin by calculating the XUV flux received at the hypothetical Earth-like exoplanet over the course of one stellar orbital period of $\sim80$ years. The orbital separation of the planet from each star and the flux received at the planet is shown in Figure \ref{fig:orbit-sim}. The flux received at the planet is dominated by $\alpha$ Cen A over the entire stellar orbital period, with the maximum flux contribution from $\alpha$ Cen B reaching only $\sim3$\% that of $\alpha$ Cen A. The total XUV flux varies between 5--11 erg s$^{-1}$ cm$^{2}$, which corresponds to approximately 1.5--3$\times$ the average XUV flux received at Earth from the Sun. It is important to note that the comparison here is to the average flux at Earth; our simulation shows that the flux received at Earth during solar maximum is comparable to that received from $\alpha$ Cen A during its minimum activity level. Previous modeling shows that Earth's atmosphere transitions to a strong hydrodynamically escaping regime if the incident EUV flux exceeds $\sim5\times$ the present day value \citep{tian_2008_xuv}; this critical flux level increases to $\sim100\times$ present day if efficient atomic line cooling is implemented \citep{nakayama_2022_xuv}. Our simulation results place the incident flux well below even the conservative limit with inefficient cooling and indicate that our hypothetical planet's atmosphere is likely to remain in a hydrostatic state.

\subsection{Atmospheric modeling}
\label{sec:atmo-model}

\begin{figure}[!ht]
    \centering
    \includegraphics[width=\linewidth]{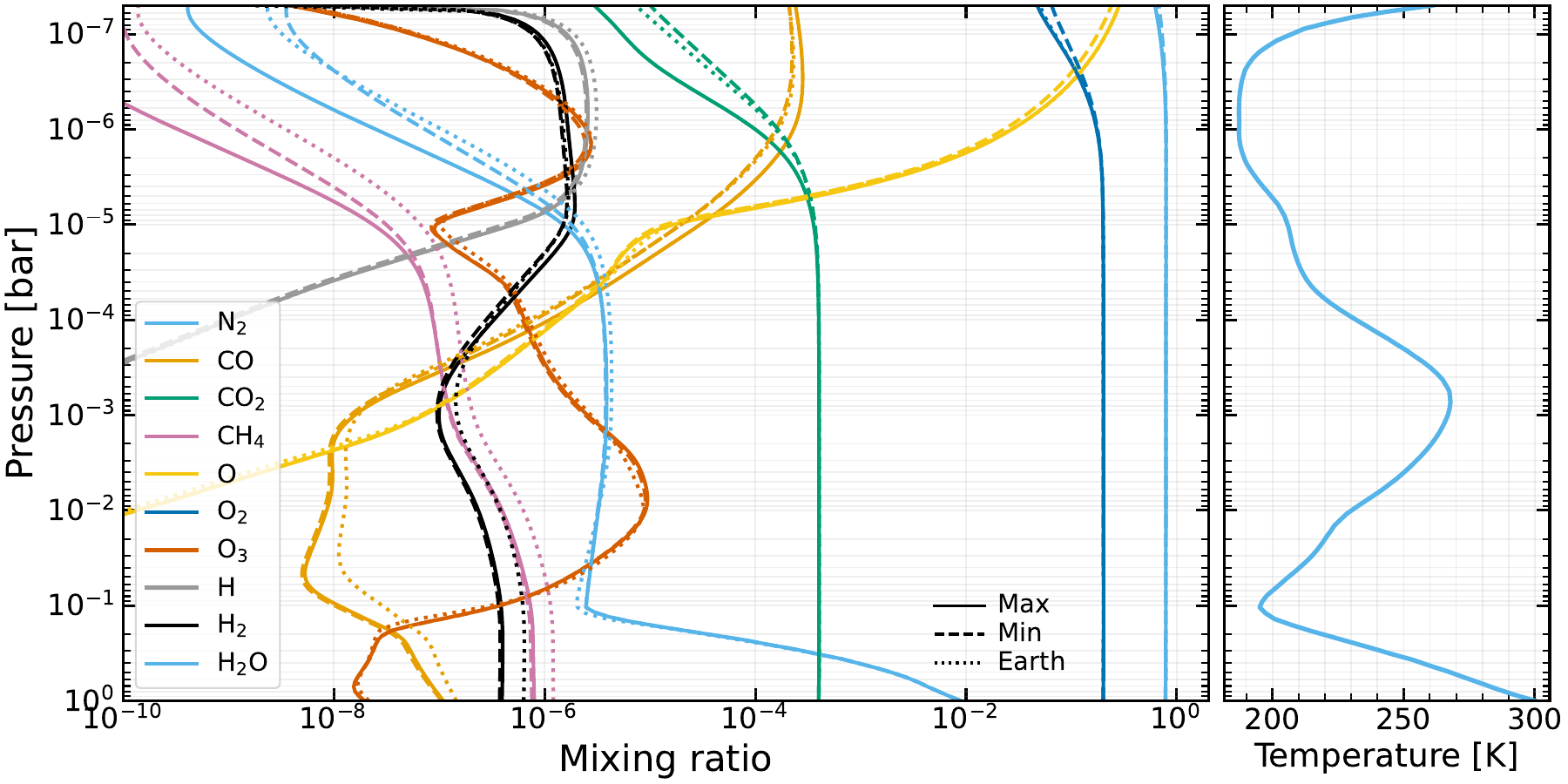}
    \caption{Atmospheric mixing ratios and temperature-pressure diagram for the simulated Earth-like terrestrial planet around $\alpha$ Cen A. Solid lines represent the case of maximum flux exposure, dashed lines the case of minimum flux exposure, and dotted lines show the profiles of Earth around the Sun retrieved from \citet{tsai_photochemically_2023}.}
    \label{fig:atm_terrestrial}
\end{figure}

We used the \texttt{VULCAN} photochemical kinetics code \citep{tsai_2021_vulcan} to simulate atmospheric chemical abundances for the hypothetical exoplanet under two flux scenarios: the ``maximum flux'' case and the ``minimum flux'' case. Maximum flux occurs when both $\alpha$ Cen A and B are at the peak of their activity cycles and $\alpha$ Cen B is at its closest approach to the planet. Minimum flux occurs when both stars are at their minimum activity level and $\alpha$ Cen B is farthest from the planet. As input to the model, we use the Earth configuration file and chemical network available from the \texttt{VULCAN} GitHub repository\footnote{\url{https://github.com/exoclime/VULCAN}}, updating the stellar parameters as necessary. This configuration includes diffusion-limited escape of H and H$_2$ from the top of the atmosphere as well as various surface sources and sinks at the bottom of the atmosphere, as described in \citet{tsai_2021_vulcan}. We also use the temperature-pressure and Eddy diffusion profiles provided by \texttt{VULCAN}. Figure \ref{fig:atm_terrestrial} shows the T-P profile and mixing ratios of our simulated planet in both the peak and minimum flux cases, as well as the mixing ratios of Earth around the Sun for comparison. The Earth profiles have been validated against the 1976 US Standard Atmosphere \citep[see][]{tsai_2021_vulcan}.

\begin{figure}[!h]
    \centering
    \includegraphics[width=\linewidth]{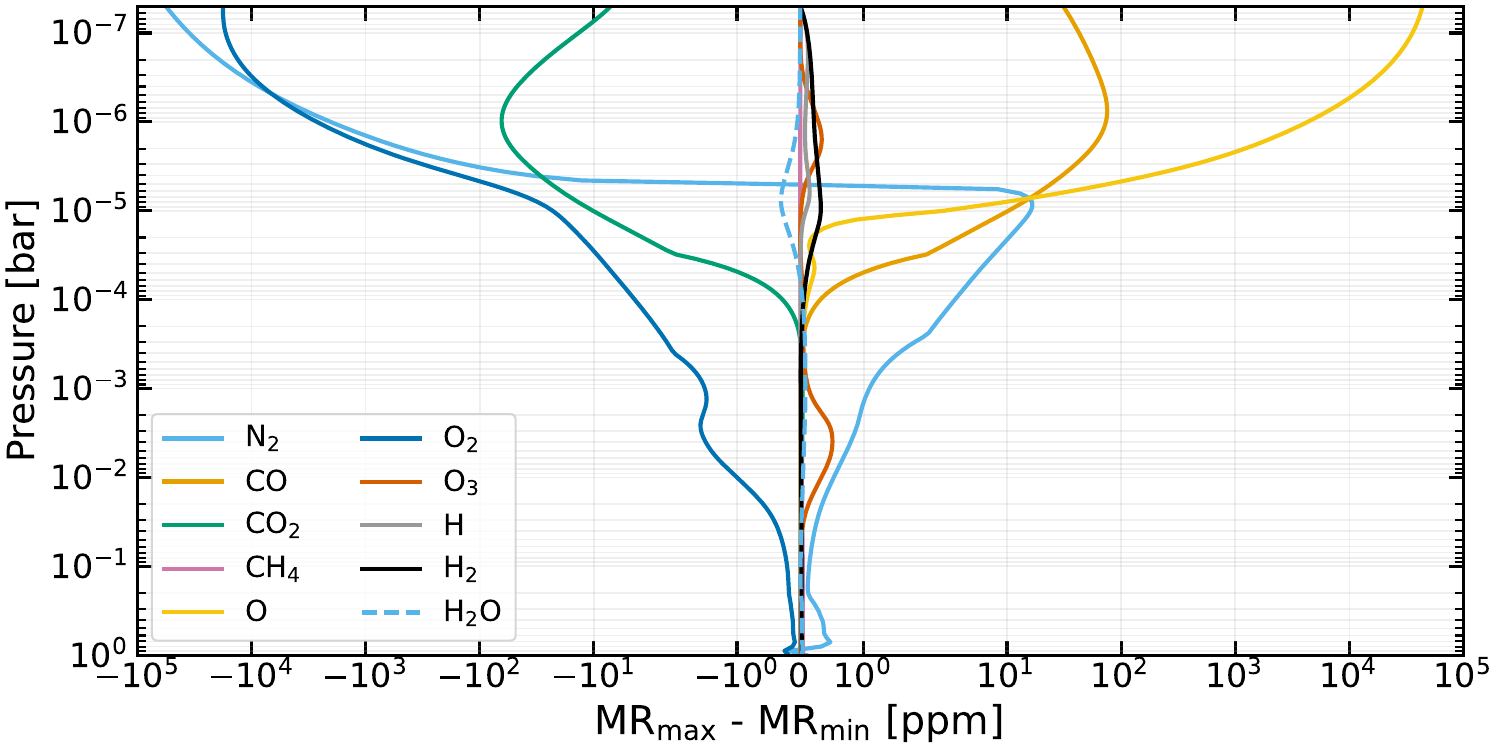}
    \caption{Difference in mixing ratio between the maximum and minimum flux scenarios for the species shown in Figure \ref{fig:atm_terrestrial}. The profile for H$_2$O is dashed because it shares the same color as N$_2$.}
    \label{fig:MR-peak-min-terrestrial}
\end{figure}

\begin{figure}[!h]
    \centering
    \includegraphics[width=\linewidth]{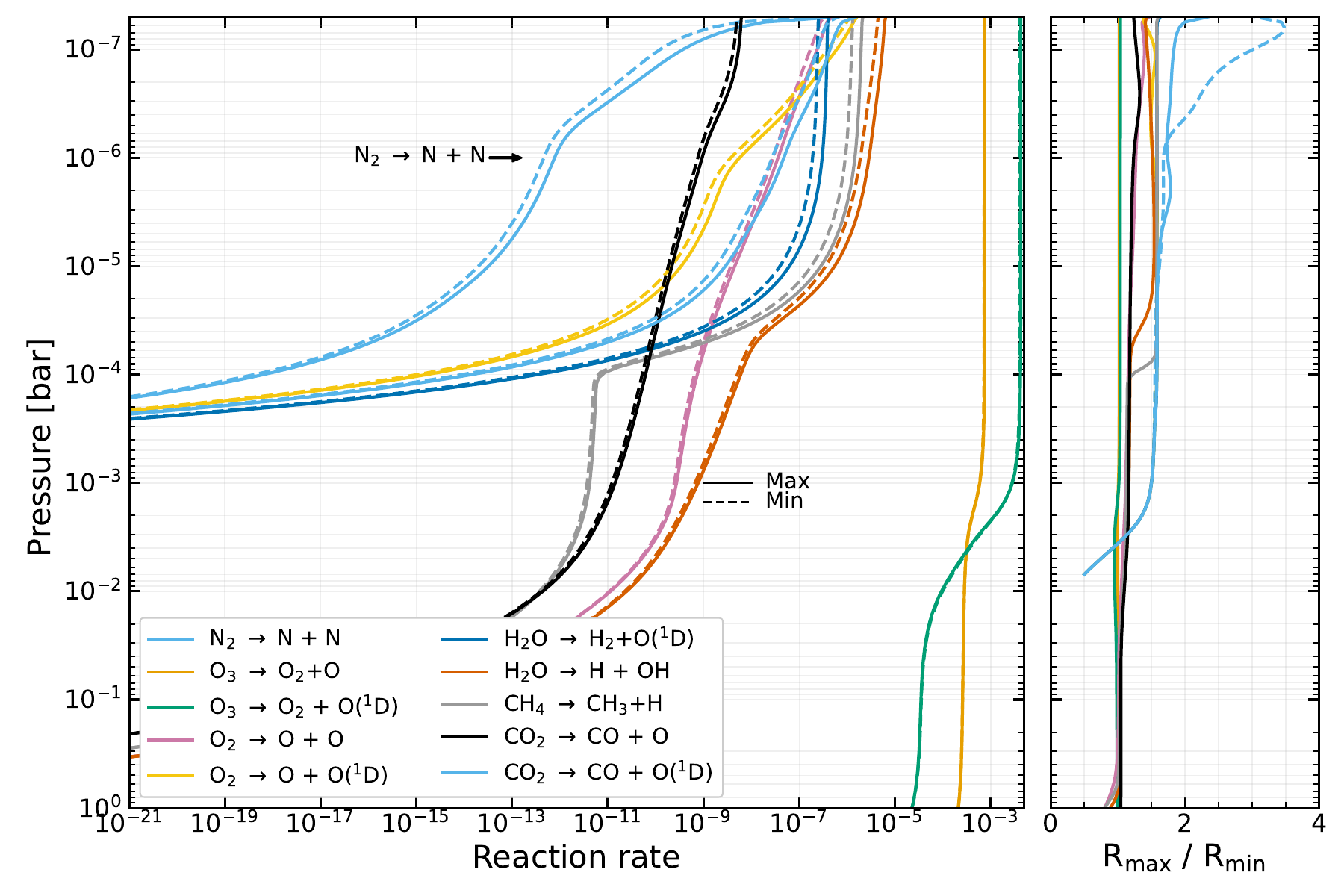}
    \caption{\textbf{Left:} Reaction rates for select dissociation reactions in the hypothetical Earth-like planet's atmosphere for both the maximum and minimum flux scenarios. \textbf{Right:} Ratio of reaction rates between the maximum and minimum flux scenarios. The reaction N$_2$ $\rightarrow$ N + N has been dashed because it shares a color with the CO$_2$ $\rightarrow$ CO + O($^1$D) reaction.}
    \label{fig:diss-rate-terrestrial}
\end{figure}

Figure \ref{fig:MR-peak-min-terrestrial} shows the difference in mixing ratio between the two flux scenarios of the species shown in Figure \ref{fig:atm_terrestrial} and Figure \ref{fig:diss-rate-terrestrial} shows the reaction rates for the dissociation of a select few N-, O-, and H-bearing species. We find no major compositional differences between any of the three models and no difference in the escape rate of H and H$_2$ at the top of the atmosphere. Between the maximum and minimum flux scenarios, we see percent-level differences in the mixing ratios of N$_2$, O$_2$, and O, and differences near 100 ppm in CO$_2$ and O$_3$. These differences are caused by the increased rate of photolysis reactions at pressures $P\lesssim0.3$ mbar, as seen in Figure \ref{fig:diss-rate-terrestrial}. Most reactions are increased by a factor of $\sim1.5$ in the maximum flux case compared to the minimum case, but the N$_2$ $\rightarrow$ N + N reaction rate reaches a peak increase of $\sim3.5\times$.

The differences in mixing ratio may be detectable by the JWST. While gas phase O$_2$ and N$_2$ are not easily detected in the N/IR \citep{ehrenreich_2006_transmission,schwieterman_2015_n2}, CO and CO$_2$ have strong absorption features at wavelengths covered by the JWST NIRSpec and MIRI, which have shown precision on the order of tens of ppm \citep{rustamkulov_analysis_2022,lustig-yaeger_transmission_2023,hu_2024_55cnc,fortune_2025_hot-rocks,bennett_2025_gj1132b}. Ultimately, however, the increased flux only introduces perturbations in mixing ratio and not noticeable differences in overall composition. Combined with the fact that the maximum flux scenario is only experienced for short periods of time and very infrequently---only a few years every $\sim19$ years---we expect the flux differences from stellar variation and from the companion star to be negligible.

\section{Summary}
\label{sec:summary}
We have presented results from the third flight of the SISTINE sounding rocket which obtained FUV spectroscopic observations of $\alpha$ Centauri A and B. The SISTINE spectra cover 980--1570 \ang\ with a resolving power of R$\sim1500$, yielding the broadest bandpass FUV spectrum of the $\alpha$ Cen system with a single observation. We used the flux-calibrated spectra, along with archival observations and model spectra, to create panchromatic SEDs from 5 \ang--1 mm. To account for the high-energy variability cycles of the stars, the SEDs employ a time- and wavelength-dependent variability model to scale the various observations, taken a many different epochs, to a common activity level.

To investigate the impact of a stellar companion on a hypothetical planet's atmosphere, we used the full SEDs to simulate the XUV flux received by an Earth-like planet orbiting $\alpha$ Cen A at 1.2 AU over one stellar orbital period, $\sim80$ years. The total XUV flux is dominated by the primary star, with the secondary contributing only up to 3\% in the most extreme case. Our atmospheric model for the hypothetical Earth-like planet around $\alpha$ Cen A suggests that the change in F/NUV flux is unlikely to cause any compositional changes in the atmosphere, with minor species having typical changes in mixing ratio on the order of ppm and major species experience differences up to a few percent.

Our simulation and modeling results imply that in the search for habitable planets around Sun-like stars, stellar binarity will not generally be an issue, as the atmospheric features will likely be dominated by the behavior of the planet-hosting star. Additionally, we note that the $\alpha$ Cen AB system is likely an extreme case when it comes to exoplanet-hosting binary systems; while the $\alpha$ Cen AB system has a semi-major axis of $\sim24$ AU, demographics studies of the distribution of confirmed planet-hosting binary systems have shown that the occurrence rate peaks for systems with semi-major axes of $\sim500$ AU, as opposed to $\sim45$ AU for field stars \citep{thebault_2025_binaries}. The flux contribution from the companion star in systems with such wide separations will be negligible.

To place our results in the context of potential HWO targets, we examined the binary stars contained in the \citet{mamajek_2024_exep-list} ExEP target list. Of the 164 targets, 48 are part of confirmed binary systems, with 9 being ``Tier A'' targets, 16 being ``Tier B'', and 23 being ``Tier C'', where Tiers A, B, and C are ranked from most-to-least accessible for imaging a hypothetical Earth-like planet with a 6-m-class telescope. Using the system distances published in the ExEP list and the semi-major axes published in the Sixth Catalog of Orbits of Visual Binary Stars \citep{hartkopf_2001_orb6}, we were able to estimate the periapse of 28 of the binary systems and compare them to that of the $\alpha$ Cen AB system. We assume, based on $1/r^2$ scaling, that any system with a periapse larger than that of $\alpha$ Cen AB will not have an impact on a potential planetary atmosphere. We found only 3 systems with periapses similar to or smaller than that of $\alpha$ Cen AB: 11 Leonis Minoris AB, 36 Ophiuchi AB, and 70 Ophiuchi AB. 11 Leonis Minoris A is a G-dwarf with $T_\mathrm{eff}=5452$ K. Its eccentric ($e=0.88$) orbit brings it within $5.18\pm0.01$ AU of its M-dwarf companion \citep{malkov_2012_binaries,soubiran_2024_gaia}. 36 Ophiuchi AB consists of two K-dwarfs both with $T_\mathrm{eff}\sim5100$ K \citep{luck_2017_abundances,luck_2018_abundances}. It is another highly eccentric system ($e=0.916$) and the stars have a periapse of $6.501\pm0.001$ AU \citep{takovinin_2017_triple}. 70 Ophiuchi AB is dynamically very similar to $\alpha$ Cen AB, with an orbital period of $P=88.34\pm0.04$ yr, eccentricity of $e=0.5005\pm0.0006$, and periapse of $a=11.58\pm0.02$ AU. The system consists of two K-dwarfs with temperatures of $T_\mathrm{eff}\sim5300$ K (A) and $T_\mathrm{eff}\sim4600$ K (B) \citep{eggenberger_2008_70oph,piccotti_2020_binaries}. All three systems mentioned here (as well as $\alpha$ Cen AB) fall into the ``Tier B'' category of \citet{mamajek_2024_exep-list} based on assumed HWO inner working angle constraints and on-sky separation of the stars.

We caution that while our results are applicable to typical Sun-like binary systems, there is still a large unexplored parameter space which may introduce complications. For example: systems with a small semi-major axis or large eccentricity, which may inhibit planet formation altogether or cause the secondary to contribute a non-negligible amount of flux; planets in very young systems, for which the host stars may be much more UV-bright than general field stars; or hosts with very active secondary stars, such as M-dwarfs, which may contribute large amounts of XUV flux through repeated flaring events. These more extreme cases should be investigated in detail individually to expand our understand of the impact of multi-star systems on habitability.

\begin{acknowledgments}
We extend our deepest gratitude to the Yolngu people of the north-eastern Arnhem land, without whom the third flight of SISTINE would not have been possible. We also thank our collaborators from the NASA Sounding Rocket Operations Contract, NASA Wallops Flight Facility, and Equatorial Launch Australia. This research was supported by NASA grants 80NSSC20K0412 and 80NSSC21K2016 (PI - K. France) to the University of Colorado Boulder. The HST, EUVE, and FUSE data used in this work can be accessed at \dataset[doi:10.17909/hvyd-mt43]{https://doi.org/10.17909/hvyd-mt43}.

\end{acknowledgments}

\vspace{5mm}
\facilities{HST, Chandra, FUSE, EUVE}

\software{\texttt{astropy} \citep{the_astropy_collaboration_astropy_2018},  
          \texttt{numpy} \citep{harris_array_2020},
          \texttt{CIAO} \citep{fruscione_ciao_2006},
          \texttt{emcee} \citep{foreman-mackey_emcee_2013},
          \texttt{VULCAN} \citep{tsai_2021_vulcan},
          \texttt{lyapy} \citep{youngblood_2022_lyapy},
          }

\bibliography{SISTINE}{}

@ARTICLE{ayres_2013_astral,
       author = {{Ayres}, T.~R.},
        title = "{Advanced Spectral Library (ASTRAL): Cool stars edition}",
      journal = {Astronomische Nachrichten},
         year = 2013,
        month = feb,
       volume = {334},
       number = {1-2},
        pages = {105},
          doi = {10.1002/asna.201211747},
       adsurl = {https://ui.adsabs.harvard.edu/abs/2013AN....334..105A}
}

@ARTICLE{ayres_2020_uv-xray,
       author = {{Ayres}, Thomas R.},
        title = "{In the Trenches of the Solar-Stellar Connection. I. Ultraviolet and X-Ray Flux-Flux Correlations across the Activity Cycles of the Sun and Alpha Centauri AB}",
      journal = {\apjs},
         year = 2020,
        month = sep,
       volume = {250},
       number = {1},
          eid = {16},
        pages = {16},
          doi = {10.3847/1538-4365/aba3c6},
       adsurl = {https://ui.adsabs.harvard.edu/abs/2020ApJS..250...16A}
}

@article{kawashima_theoretical_2018,
	title = {Theoretical {Transmission} {Spectra} of {Exoplanet} {Atmospheres} with {Hydrocarbon} {Haze}: {Effect} of {Creation}, {Growth}, and {Settling} of {Haze} {Particles}. {I}. {Model} {Description} and {First} {Results}},
	volume = {853},
	issn = {0004-637X, 1538-4357},
	shorttitle = {Theoretical {Transmission} {Spectra} of {Exoplanet} {Atmospheres} with {Hydrocarbon} {Haze}},
	url = {https://iopscience.iop.org/article/10.3847/1538-4357/aaa0c5},
	doi = {10.3847/1538-4357/aaa0c5},
	language = {en},
	number = {1},
	urldate = {2024-06-28},
	journal = {The Astrophysical Journal},
	author = {Kawashima, Yui and Ikoma, Masahiro},
	month = jan,
	year = {2018},
	pages = {7},
}

@article{he_laboratory_2018,
	title = {Laboratory {Simulations} of {Haze} {Formation} in the {Atmospheres} of {Super}-{Earths} and {Mini}-{Neptunes}: {Particle} {Color} and {Size} {Distribution}},
	volume = {856},
	issn = {2041-8205, 2041-8213},
	shorttitle = {Laboratory {Simulations} of {Haze} {Formation} in the {Atmospheres} of {Super}-{Earths} and {Mini}-{Neptunes}},
	url = {https://iopscience.iop.org/article/10.3847/2041-8213/aab42b},
	doi = {10.3847/2041-8213/aab42b},
	language = {en},
	number = {1},
	urldate = {2024-06-28},
	journal = {The Astrophysical Journal Letters},
	author = {He, Chao and Hörst, Sarah M. and Lewis, Nikole K. and Yu, Xinting and Moses, Julianne I. and Kempton, Eliza M.-R. and McGuiggan, Patricia and Morley, Caroline V. and Valenti, Jeff A. and Vuitton, Véronique},
	month = mar,
	year = {2018},
	pages = {L3},
}

@article{the_astropy_collaboration_astropy_2018,
	title = {The {Astropy} {Project}: {Building} an {Open}-science {Project} and {Status} of the v2.0 {Core} {Package} $^{\textrm{*}}$},
	volume = {156},
	issn = {0004-6256, 1538-3881},
	shorttitle = {The {Astropy} {Project}},
	url = {https://iopscience.iop.org/article/10.3847/1538-3881/aabc4f},
	doi = {10.3847/1538-3881/aabc4f},
	language = {en},
	number = {3},
	urldate = {2024-06-28},
	journal = {The Astronomical Journal},
	author = {{The Astropy Collaboration} and Price-Whelan, A. M. and Sipőcz, B. M. and Günther, H. M. and Lim, P. L. and Crawford, S. M. and Conseil, S. and Shupe, D. L. and Craig, M. W. and Dencheva, N. and Ginsburg, A. and VanderPlas, J. T. and Bradley, L. D. and Pérez-Suárez, D. and De Val-Borro, M. and {(Primary Paper Contributors)} and Aldcroft, T. L. and Cruz, K. L. and Robitaille, T. P. and Tollerud, E. J. and {(Astropy Coordination Committee)} and Ardelean, C. and Babej, T. and Bach, Y. P. and Bachetti, M. and Bakanov, A. V. and Bamford, S. P. and Barentsen, G. and Barmby, P. and Baumbach, A. and Berry, K. L. and Biscani, F. and Boquien, M. and Bostroem, K. A. and Bouma, L. G. and Brammer, G. B. and Bray, E. M. and Breytenbach, H. and Buddelmeijer, H. and Burke, D. J. and Calderone, G. and Rodríguez, J. L. Cano and Cara, M. and Cardoso, J. V. M. and Cheedella, S. and Copin, Y. and Corrales, L. and Crichton, D. and D’Avella, D. and Deil, C. and Depagne, É. and Dietrich, J. P. and Donath, A. and Droettboom, M. and Earl, N. and Erben, T. and Fabbro, S. and Ferreira, L. A. and Finethy, T. and Fox, R. T. and Garrison, L. H. and Gibbons, S. L. J. and Goldstein, D. A. and Gommers, R. and Greco, J. P. and Greenfield, P. and Groener, A. M. and Grollier, F. and Hagen, A. and Hirst, P. and Homeier, D. and Horton, A. J. and Hosseinzadeh, G. and Hu, L. and Hunkeler, J. S. and Ivezić, Ž. and Jain, A. and Jenness, T. and Kanarek, G. and Kendrew, S. and Kern, N. S. and Kerzendorf, W. E. and Khvalko, A. and King, J. and Kirkby, D. and Kulkarni, A. M. and Kumar, A. and Lee, A. and Lenz, D. and Littlefair, S. P. and Ma, Z. and Macleod, D. M. and Mastropietro, M. and McCully, C. and Montagnac, S. and Morris, B. M. and Mueller, M. and Mumford, S. J. and Muna, D. and Murphy, N. A. and Nelson, S. and Nguyen, G. H. and Ninan, J. P. and Nöthe, M. and Ogaz, S. and Oh, S. and Parejko, J. K. and Parley, N. and Pascual, S. and Patil, R. and Patil, A. A. and Plunkett, A. L. and Prochaska, J. X. and Rastogi, T. and Janga, V. Reddy and Sabater, J. and Sakurikar, P. and Seifert, M. and Sherbert, L. E. and Sherwood-Taylor, H. and Shih, A. Y. and Sick, J. and Silbiger, M. T. and Singanamalla, S. and Singer, L. P. and Sladen, P. H. and Sooley, K. A. and Sornarajah, S. and Streicher, O. and Teuben, P. and Thomas, S. W. and Tremblay, G. R. and Turner, J. E. H. and Terrón, V. and Kerkwijk, M. H. Van and De La Vega, A. and Watkins, L. L. and Weaver, B. A. and Whitmore, J. B. and Woillez, J. and Zabalza, V. and {(Astropy Contributors)}},
	month = sep,
	year = {2018},
	pages = {123},
}

@article{loyd_muscles_2018,
	title = {The {MUSCLES} {Treasury} {Survey}. {V}. {FUV} {Flares} on {Active} and {Inactive} {M} {Dwarfs}*†‡},
	volume = {867},
	issn = {0004-637X, 1538-4357},
	url = {https://iopscience.iop.org/article/10.3847/1538-4357/aae2bd},
	doi = {10.3847/1538-4357/aae2bd},
	language = {en},
	number = {1},
	urldate = {2024-06-28},
	journal = {The Astrophysical Journal},
	author = {Loyd, R. O. Parke and France, Kevin and Youngblood, Allison and Schneider, Christian and Brown, Alexander and Hu, Renyu and Segura, Antígona and Linsky, Jeffrey and Redfield, Seth and Tian, Feng and Rugheimer, Sarah and Miguel, Yamila and Froning, Cynthia S.},
	month = nov,
	year = {2018},
	pages = {71},
}

@article{froning_hot_2019,
	title = {A {Hot} {Ultraviolet} {Flare} on the {M} {Dwarf} {Star} {GJ} 674},
	volume = {871},
	issn = {2041-8205},
	url = {https://dx.doi.org/10.3847/2041-8213/aaffcd},
	doi = {10.3847/2041-8213/aaffcd},
	language = {en},
	number = {2},
	urldate = {2024-06-28},
	journal = {The Astrophysical Journal Letters},
	author = {Froning, Cynthia S. and Kowalski, Adam and France, Kevin and Loyd, R. O. Parke and Schneider, P. Christian and Youngblood, Allison and Wilson, David and Brown, Alexander and Berta-Thompson, Zachory and Pineda, J. Sebastian and Linsky, Jeffrey and Rugheimer, Sarah and Miguel, Yamila},
	month = jan,
	year = {2019},
	note = {Publisher: The American Astronomical Society},
	pages = {L26},
}

@article{fortney_hot_2021,
	title = {Hot {Jupiters}: {Origins}, {Structure}, {Atmospheres}},
	volume = {126},
	issn = {2169-9097, 2169-9100},
	shorttitle = {Hot {Jupiters}},
	url = {https://agupubs.onlinelibrary.wiley.com/doi/10.1029/2020JE006629},
	doi = {10.1029/2020JE006629},
	language = {en},
	number = {3},
	urldate = {2024-06-28},
	journal = {Journal of Geophysical Research: Planets},
	author = {Fortney, Jonathan J. and Dawson, Rebekah I. and Komacek, Thaddeus D.},
	month = mar,
	year = {2021},
	pages = {e2020JE006629},
}

@article{youngblood_intrinsic_2022,
	title = {Intrinsic {Lyα} {Profiles} of {High}-velocity {G}, {K}, and {M} {Dwarfs}},
	volume = {926},
	issn = {0004-637X},
	url = {https://dx.doi.org/10.3847/1538-4357/ac4711},
	doi = {10.3847/1538-4357/ac4711},
	language = {en},
	number = {2},
	urldate = {2024-06-28},
	journal = {The Astrophysical Journal},
	author = {Youngblood, Allison and Pineda, J. Sebastian and Ayres, Thomas and France, Kevin and Linsky, Jeffrey L. and Wood, Brian E. and Redfield, Seth and Schlieder, Joshua E.},
	month = feb,
	year = {2022},
	note = {Publisher: The American Astronomical Society},
	pages = {129},
}

@article{tsai_photochemically_2023,
	title = {Photochemically produced {SO2} in the atmosphere of {WASP}-39b},
	volume = {617},
	issn = {1476-4687},
	doi = {10.1038/s41586-023-05902-2},
	language = {eng},
	number = {7961},
	journal = {Nature},
	author = {Tsai, Shang-Min and Lee, Elspeth K. H. and Powell, Diana and Gao, Peter and Zhang, Xi and Moses, Julianne and Hébrard, Eric and Venot, Olivia and Parmentier, Vivien and Jordan, Sean and Hu, Renyu and Alam, Munazza K. and Alderson, Lili and Batalha, Natalie M. and Bean, Jacob L. and Benneke, Björn and Bierson, Carver J. and Brady, Ryan P. and Carone, Ludmila and Carter, Aarynn L. and Chubb, Katy L. and Inglis, Julie and Leconte, Jérémy and Line, Michael and López-Morales, Mercedes and Miguel, Yamila and Molaverdikhani, Karan and Rustamkulov, Zafar and Sing, David K. and Stevenson, Kevin B. and Wakeford, Hannah R. and Yang, Jeehyun and Aggarwal, Keshav and Baeyens, Robin and Barat, Saugata and de Val-Borro, Miguel and Daylan, Tansu and Fortney, Jonathan J. and France, Kevin and Goyal, Jayesh M. and Grant, David and Kirk, James and Kreidberg, Laura and Louca, Amy and Moran, Sarah E. and Mukherjee, Sagnick and Nasedkin, Evert and Ohno, Kazumasa and Rackham, Benjamin V. and Redfield, Seth and Taylor, Jake and Tremblin, Pascal and Visscher, Channon and Wallack, Nicole L. and Welbanks, Luis and Youngblood, Allison and Ahrer, Eva-Maria and Batalha, Natasha E. and Behr, Patrick and Berta-Thompson, Zachory K. and Blecic, Jasmina and Casewell, S. L. and Crossfield, Ian J. M. and Crouzet, Nicolas and Cubillos, Patricio E. and Decin, Leen and Désert, Jean-Michel and Feinstein, Adina D. and Gibson, Neale P. and Harrington, Joseph and Heng, Kevin and Henning, Thomas and Kempton, Eliza M.-R. and Krick, Jessica and Lagage, Pierre-Olivier and Lendl, Monika and Lothringer, Joshua D. and Mansfield, Megan and Mayne, N. J. and Mikal-Evans, Thomas and Palle, Enric and Schlawin, Everett and Shorttle, Oliver and Wheatley, Peter J. and Yurchenko, Sergei N.},
	month = may,
	year = {2023},
	pmid = {37100917},
	pmcid = {PMC10191860},
	pages = {483--487},
}

@article{linsky_intrinsic_2014,
	title = {The {Intrinsic} {Extreme} {Ultraviolet} {Fluxes} of {F5} {V} to {M5} {V} {Stars}},
	volume = {780},
	issn = {0004-637X},
	url = {https://dx.doi.org/10.1088/0004-637X/780/1/61},
	doi = {10.1088/0004-637X/780/1/61},
	language = {en},
	number = {1},
	urldate = {2024-06-28},
	journal = {The Astrophysical Journal},
	author = {Linsky, Jeffrey L. and Fontenla, Juan and France, Kevin},
	month = jan,
	year = {2014},
	note = {Publisher: The American Astronomical Society},
	pages = {61},
}

@article{france_ultraviolet_2013,
	title = {The {Ultraviolet} {Radiation} {Envrionment} {Around} {M} {Dwarf} {Exoplanet} {Host} {Stars}},
	volume = {763},
	issn = {0004-637X},
	url = {https://dx.doi.org/10.1088/0004-637X/763/2/149},
	doi = {10.1088/0004-637X/763/2/149},
	language = {en},
	number = {2},
	urldate = {2024-06-28},
	journal = {The Astrophysical Journal},
	author = {France, Kevin and Froning, Cynthia S. and Linsky, Jeffrey L. and Roberge, Aki and Stocke, John T. and Tian, Feng and Bushinsky, Rachel and Désert, Jean-Michel and Mauas, Pablo and Vieytes, Mariela and Walkowicz, Lucianne M.},
	month = jan,
	year = {2013},
	note = {Publisher: The American Astronomical Society},
	pages = {149},
}

@article{hu_photochemistry_2012,
	title = {Photochemistry in {Terrestrial} {Exoplanet} {Atmospherse} {I}. {Photochemistry} {Model} and {Benchmark} {Cases}},
	volume = {761},
	issn = {0004-637X},
	url = {https://dx.doi.org/10.1088/0004-637X/761/2/166},
	doi = {10.1088/0004-637X/761/2/166},
	language = {en},
	number = {2},
	urldate = {2024-06-28},
	journal = {The Astrophysical Journal},
	author = {Hu, Renyu and Seager, Sara and Bains, William},
	month = dec,
	year = {2012},
	note = {Publisher: The American Astronomical Society},
	pages = {166},
}

@article{miguel_effect_2015,
	title = {The effect of {Lyman} α radiation on mini-{Neptune} atmospheres around {M} stars: application to {GJ} 436b},
	volume = {446},
	issn = {0035-8711},
	shorttitle = {The effect of {Lyman} α radiation on mini-{Neptune} atmospheres around {M} stars},
	url = {https://doi.org/10.1093/mnras/stu2107},
	doi = {10.1093/mnras/stu2107},
	number = {1},
	urldate = {2024-06-28},
	journal = {Monthly Notices of the Royal Astronomical Society},
	author = {Miguel, Yamila and Kaltenegger, Lisa and Linsky, Jeffrey L. and Rugheimer, Sarah},
	month = jan,
	year = {2015},
	pages = {345--353},
}

@article{loyd_muscles_2016,
	title = {The {MUSCLES} {Treasury} {Survey} {III}. {X}-ray to {Infrared} {Spectra} of 11 {M} and {K} {Stars} {Hosting} {Planets}},
	volume = {824},
	issn = {0004-637X},
	url = {https://dx.doi.org/10.3847/0004-637X/824/2/102},
	doi = {10.3847/0004-637X/824/2/102},
	language = {en},
	number = {2},
	urldate = {2024-06-28},
	journal = {The Astrophysical Journal},
	author = {Loyd, R. O. P. and France, Kevin and Youngblood, Allison and Schneider, Christian and Brown, Alexander and Hu, Renyu and Linsky, Jeffrey and Froning, Cynthia S. and Redfield, Seth and Rugheimer, Sarah and Tian, Feng},
	month = jun,
	year = {2016},
	note = {Publisher: The American Astronomical Society},
	pages = {102},
}

@article{murray-clay_atmospheric_2009,
	title = {Atmospheric {Escape} from {Hot} {Jupiters}},
	volume = {693},
	issn = {0004-637X},
	url = {https://dx.doi.org/10.1088/0004-637X/693/1/23},
	doi = {10.1088/0004-637X/693/1/23},
	language = {en},
	number = {1},
	urldate = {2024-06-28},
	journal = {The Astrophysical Journal},
	author = {Murray-Clay, Ruth A. and Chiang, Eugene I. and Murray, Norman},
	month = feb,
	year = {2009},
	note = {Publisher: The American Astronomical Society},
	pages = {23},
}

@article{vidal-madjar_detection_2004,
	title = {Detection of {Oxygen} and {Carbon} in the {Hydrodynamically} {Escaping} {Atmosphere} of the {Extrasolar} {Planet} {HD} 209458b},
	volume = {604},
	issn = {0004-637X},
	url = {https://iopscience.iop.org/article/10.1086/383347/meta},
	doi = {10.1086/383347},
	language = {en},
	number = {1},
	urldate = {2024-06-28},
	journal = {The Astrophysical Journal},
	author = {Vidal-Madjar, A. and Désert, J.-M. and Etangs, A. Lecavelier des and Hébrard, G. and Ballester, G. E. and Ehrenreich, D. and Ferlet, R. and McConnell, J. C. and Mayor, M. and Parkinson, C. D.},
	month = mar,
	year = {2004},
	note = {Publisher: IOP Publishing},
	pages = {L69},
}

@ARTICLE{rabl-dvorak_1998_binaries,
       author = {{Rabl}, G. and {Dvorak}, R.},
        title = "{Satellite-type planetary orbits in double stars : a numerical approach.}",
      journal = {\aap},
         year = 1988,
        month = feb,
       volume = {191},
        pages = {385-391},
       adsurl = {https://ui.adsabs.harvard.edu/abs/1988A&A...191..385R}
}

@ARTICLE{nakayama_2022_xuv,
       author = {{Nakayama}, Akifumi and {Ikoma}, Masahiro and {Terada}, Naoki},
        title = "{Survival of Terrestrial N$_{2}$-O$_{2}$ Atmospheres in Violent XUV Environments through Efficient Atomic Line Radiative Cooling}",
      journal = {\apj},
         year = 2022,
        month = oct,
       volume = {937},
       number = {2},
          eid = {72},
        pages = {72},
          doi = {10.3847/1538-4357/ac86ca},
archivePrefix = {arXiv},
       eprint = {2210.01460},
 primaryClass = {astro-ph.EP},
       adsurl = {https://ui.adsabs.harvard.edu/abs/2022ApJ...937...72N}
}

@ARTICLE{zhao_2018_detectability,
       author = {{Zhao}, Lily and {Fischer}, Debra A. and {Brewer}, John and {Giguere}, Matt and {Rojas-Ayala}, B{\'a}rbara},
        title = "{Planet Detectability in the Alpha Centauri System}",
      journal = {\aj},
         year = 2018,
        month = jan,
       volume = {155},
       number = {1},
          eid = {24},
        pages = {24},
          doi = {10.3847/1538-3881/aa9bea},
archivePrefix = {arXiv},
       eprint = {1711.06320},
 primaryClass = {astro-ph.EP},
       adsurl = {https://ui.adsabs.harvard.edu/abs/2018AJ....155...24Z}
}

@ARTICLE{fortune_2025_hot-rocks,
       author = {{Fortune}, Mark and {Gibson}, Neale P. and {Diamond-Lowe}, Hannah and {Mendon{\c{c}}a}, Jo{\~a}o M. and {Gressier}, Am{\'e}lie and {Kitzmann}, Daniel and {Allen}, Natalie H. and {August}, Prune C. and {Ih}, Jegug and {Meier Vald{\'e}s}, Erik and et al.},
        title = "{Hot Rocks Survey: III. A deep eclipse for LHS 1140c and a new Gaussian process method to account for correlated noise in individual pixels}",
      journal = {\aap},
         year = 2025,
        month = sep,
       volume = {701},
          eid = {A25},
        pages = {A25},
          doi = {10.1051/0004-6361/202554198},
archivePrefix = {arXiv},
       eprint = {2505.22186},
 primaryClass = {astro-ph.EP},
       adsurl = {https://ui.adsabs.harvard.edu/abs/2025A&A...701A..25F}
}

@ARTICLE{hu_2024_55cnc,
       author = {{Hu}, Renyu and {Bello-Arufe}, Aaron and {Zhang}, Michael and {Paragas}, Kimberly and {Zilinskas}, Mantas and {van Buchem}, Christiaan and {Bess}, Michael and {Patel}, Jayshil and {Ito}, Yuichi and {Damiano}, Mario and et al.},
        title = "{A secondary atmosphere on the rocky exoplanet 55 Cancri e}",
      journal = {\nat},
         year = 2024,
        month = jun,
       volume = {630},
       number = {8017},
        pages = {609-612},
          doi = {10.1038/s41586-024-07432-x},
archivePrefix = {arXiv},
       eprint = {2405.04744},
 primaryClass = {astro-ph.EP},
       adsurl = {https://ui.adsabs.harvard.edu/abs/2024Natur.630..609H}
}

@ARTICLE{bennett_2025_gj1132b,
       author = {{Bennett}, Katherine A. and {MacDonald}, Ryan J. and {Peacock}, Sarah and {Perez}, Junellie and {May}, E.~M. and {Moran}, Sarah E. and {Alderson}, Lili and {Lustig-Yaeger}, Jacob and {Wakeford}, Hannah R. and {Sing}, David K. and et al.},
        title = "{Additional JWST/NIRSpec Transits of the Rocky M Dwarf Exoplanet GJ 1132 b Reveal a Featureless Spectrum}",
      journal = {\aj},
         year = 2025,
        month = oct,
       volume = {170},
       number = {4},
          eid = {205},
        pages = {205},
          doi = {10.3847/1538-3881/adf198},
archivePrefix = {arXiv},
       eprint = {2508.10579},
 primaryClass = {astro-ph.EP},
       adsurl = {https://ui.adsabs.harvard.edu/abs/2025AJ....170..205B}
}

@ARTICLE{schwieterman_2015_n2,
       author = {{Schwieterman}, Edward W. and {Robinson}, Tyler D. and {Meadows}, Victoria S. and {Misra}, Amit and {Domagal-Goldman}, Shawn},
        title = "{Detecting and Constraining N$_{2}$ Abundances in Planetary Atmospheres Using Collisional Pairs}",
      journal = {\apj},
         year = 2015,
        month = sep,
       volume = {810},
       number = {1},
          eid = {57},
        pages = {57},
          doi = {10.1088/0004-637X/810/1/57},
archivePrefix = {arXiv},
       eprint = {1507.07945},
 primaryClass = {astro-ph.EP},
       adsurl = {https://ui.adsabs.harvard.edu/abs/2015ApJ...810...57S}
}

@ARTICLE{ehrenreich_2006_transmission,
       author = {{Ehrenreich}, D. and {Tinetti}, G. and {Lecavelier Des Etangs}, A. and {Vidal-Madjar}, A. and {Selsis}, F.},
        title = "{The transmission spectrum of Earth-size transiting planets}",
      journal = {\aap},
         year = 2006,
        month = mar,
       volume = {448},
       number = {1},
        pages = {379-393},
          doi = {10.1051/0004-6361:20053861},
archivePrefix = {arXiv},
       eprint = {astro-ph/0510215},
 primaryClass = {astro-ph},
       adsurl = {https://ui.adsabs.harvard.edu/abs/2006A&A...448..379E}
}

@ARTICLE{drake_1997_euve,
       author = {{Drake}, Jeremy J. and {Laming}, J. Martin and {Widing}, Kenneth G.},
        title = "{Stellar Coronal Abundances. V. Evidence for the First Ionization Potential Effect in {\ensuremath{\alpha}} Centauri}",
      journal = {\apj},
         year = 1997,
        month = mar,
       volume = {478},
       number = {1},
        pages = {403-416},
          doi = {10.1086/303755},
       adsurl = {https://ui.adsabs.harvard.edu/abs/1997ApJ...478..403D}
}

@ARTICLE{mamajek_2024_exep-list,
       author = {{Mamajek}, Eric and {Stapelfeldt}, Karl},
        title = "{NASA Exoplanet Exploration Program (ExEP) Mission Star List for the Habitable Worlds Observatory (2023)}",
      journal = {arXiv e-prints},
         year = 2024,
        month = feb,
          eid = {arXiv:2402.12414},
        pages = {arXiv:2402.12414},
          doi = {10.48550/arXiv.2402.12414},
archivePrefix = {arXiv},
       eprint = {2402.12414},
 primaryClass = {astro-ph.IM},
       adsurl = {https://ui.adsabs.harvard.edu/abs/2024arXiv240212414M}
}

@software{youngblood_2022_lyapy,
       author = {{Youngblood}, Allison and {Newton}, Elisabeth R.},
        title = "{allisony/lyapy: First release created for citation purposes in the literature}",
         year = 2022,
        month = aug,
          eid = {10.5281/zenodo.6949067},
          doi = {10.5281/zenodo.6949067},
      version = {v1.0.0},
    publisher = {Zenodo},
       adsurl = {https://ui.adsabs.harvard.edu/abs/2022zndo...6949067Y}
}

@ARTICLE{tsai_2021_vulcan,
       author = {{Tsai}, Shang-Min and {Malik}, Matej and {Kitzmann}, Daniel and {Lyons}, James R. and {Fateev}, Alexander and {Lee}, Elspeth and {Heng}, Kevin},
        title = "{A Comparative Study of Atmospheric Chemistry with VULCAN}",
      journal = {\apj},
         year = 2021,
        month = dec,
       volume = {923},
       number = {2},
          eid = {264},
        pages = {264},
          doi = {10.3847/1538-4357/ac29bc},
archivePrefix = {arXiv},
       eprint = {2108.01790},
 primaryClass = {astro-ph.EP},
       adsurl = {https://ui.adsabs.harvard.edu/abs/2021ApJ...923..264T}
}

@ARTICLE{tuchow_2024_hpic,
       author = {{Tuchow}, Noah W. and {Stark}, Christopher C. and {Mamajek}, Eric},
        title = "{HPIC: The Habitable Worlds Observatory Preliminary Input Catalog}",
      journal = {\aj},
         year = 2024,
        month = mar,
       volume = {167},
       number = {3},
          eid = {139},
        pages = {139},
          doi = {10.3847/1538-3881/ad25ec},
       adsurl = {https://ui.adsabs.harvard.edu/abs/2024AJ....167..139T}
}

@ARTICLE{kraus_2016_impact-of-multiplicity,
       author = {{Kraus}, Adam L. and {Ireland}, Michael J. and {Huber}, Daniel and {Mann}, Andrew W. and {Dupuy}, Trent J.},
        title = "{The Impact of Stellar Multiplicity on Planetary Systems. I. The Ruinous Influence of Close Binary Companions}",
      journal = {\aj},
         year = 2016,
        month = jul,
       volume = {152},
       number = {1},
          eid = {8},
        pages = {8},
          doi = {10.3847/0004-6256/152/1/8},
archivePrefix = {arXiv},
       eprint = {1604.05744},
 primaryClass = {astro-ph.EP},
       adsurl = {https://ui.adsabs.harvard.edu/abs/2016AJ....152....8K}
}

@ARTICLE{moe_2021_binary-statistics,
       author = {{Moe}, Maxwell and {Kratter}, Kaitlin M.},
        title = "{Impact of binary stars on planet statistics - I. Planet occurrence rates and trends with stellar mass}",
      journal = {\mnras},
         year = 2021,
        month = nov,
       volume = {507},
       number = {3},
        pages = {3593-3611},
          doi = {10.1093/mnras/stab2328},
archivePrefix = {arXiv},
       eprint = {1912.01699},
 primaryClass = {astro-ph.EP},
       adsurl = {https://ui.adsabs.harvard.edu/abs/2021MNRAS.507.3593M}
}

@ARTICLE{quarles_2020_binary,
       author = {{Quarles}, Billy and {Li}, Gongjie and {Kostov}, Veselin and {Haghighipour}, Nader},
        title = "{Orbital Stability of Circumstellar Planets in Binary Systems}",
      journal = {\aj},
         year = 2020,
        month = mar,
       volume = {159},
       number = {3},
          eid = {80},
        pages = {80},
          doi = {10.3847/1538-3881/ab64fa},
archivePrefix = {arXiv},
       eprint = {1912.11019},
 primaryClass = {astro-ph.EP},
       adsurl = {https://ui.adsabs.harvard.edu/abs/2020AJ....159...80Q}
}

@ARTICLE{holman_1999_binary,
       author = {{Holman}, Matthew J. and {Wiegert}, Paul A.},
        title = "{Long-Term Stability of Planets in Binary Systems}",
      journal = {\aj},
         year = 1999,
        month = jan,
       volume = {117},
       number = {1},
        pages = {621-628},
          doi = {10.1086/300695},
archivePrefix = {arXiv},
       eprint = {astro-ph/9809315},
 primaryClass = {astro-ph},
       adsurl = {https://ui.adsabs.harvard.edu/abs/1999AJ....117..621H}
}

@ARTICLE{wiegert_1997_alpha-cen,
       author = {{Wiegert}, Paul A. and {Holman}, Matt J.},
        title = "{The Stability of Planets in the Alpha Centauri System}",
      journal = {\aj},
         year = 1997,
        month = apr,
       volume = {113},
        pages = {1445-1450},
          doi = {10.1086/118360},
archivePrefix = {arXiv},
       eprint = {astro-ph/9609106},
 primaryClass = {astro-ph},
       adsurl = {https://ui.adsabs.harvard.edu/abs/1997AJ....113.1445W}
}

@ARTICLE{linsky_1996_ism,
       author = {{Linsky}, Jeffrey L. and {Wood}, Brian E.},
        title = "{The {\ensuremath{\alpha}} Centauri line of sight: D/H ratio, physical properties of local interstellar gas, and measurement of heated hydrogen (the ``hydrogen wall'') near the heliopause.}",
      journal = {\apj},
         year = 1996,
        month = may,
       volume = {463},
        pages = {254-270},
          doi = {10.1086/177238},
       adsurl = {https://ui.adsabs.harvard.edu/abs/1996ApJ...463..254L}
}

@ARTICLE{ayres_2014_cycles,
       author = {{Ayres}, Thomas R.},
        title = "{The Ups and Downs of {\ensuremath{\alpha}} Centauri}",
      journal = {\aj},
         year = 2014,
        month = mar,
       volume = {147},
       number = {3},
          eid = {59},
        pages = {59},
          doi = {10.1088/0004-6256/147/3/59},
archivePrefix = {arXiv},
       eprint = {1401.0847},
 primaryClass = {astro-ph.SR},
       adsurl = {https://ui.adsabs.harvard.edu/abs/2014AJ....147...59A}
}

@ARTICLE{france_2025_euv,
       author = {{France}, Kevin and {Duvvuri}, Girish and {Froning}, Cynthia S. and {Brown}, Alexander and {Schneider}, P. Christian and {Pineda}, J. Sebastian and {Wilson}, David and {Youngblood}, Allison and {Airapetian}, Vladimir S. and {Namekata}, Kosuke and {Notsu}, Yuta and {Sextro}, Tristen},
        title = "{A Semiempirical Estimate of Solar Extreme-ultraviolet Evolution from 10 Myr to 10 Gyr}",
      journal = {\aj},
         year = 2025,
        month = sep,
       volume = {170},
       number = {3},
          eid = {159},
        pages = {159},
          doi = {10.3847/1538-3881/adefdf},
archivePrefix = {arXiv},
       eprint = {2507.15953},
 primaryClass = {astro-ph.SR},
       adsurl = {https://ui.adsabs.harvard.edu/abs/2025AJ....170..159F}
}

@ARTICLE{france_2018_fuv,
       author = {{France}, Kevin and {Arulanantham}, Nicole and {Fossati}, Luca and {Lanza}, Antonino F. and {Loyd}, R.~O. Parke and {Redfield}, Seth and {Schneider}, P. Christian},
        title = "{Far-ultraviolet Activity Levels of F, G, K, and M Dwarf Exoplanet Host Stars}",
      journal = {\apjs},
         year = 2018,
        month = nov,
       volume = {239},
       number = {1},
          eid = {16},
        pages = {16},
          doi = {10.3847/1538-4365/aae1a3},
archivePrefix = {arXiv},
       eprint = {1809.07342},
 primaryClass = {astro-ph.SR},
       adsurl = {https://ui.adsabs.harvard.edu/abs/2018ApJS..239...16F}
}

@ARTICLE{youngblood_2016_muscles,
       author = {{Youngblood}, Allison and {France}, Kevin and {Loyd}, R.~O. Parke and {Linsky}, Jeffrey L. and {Redfield}, Seth and {Schneider}, P. Christian and {Wood}, Brian E. and {Brown}, Alexander and {Froning}, Cynthia and {Miguel}, Yamila and {Rugheimer}, Sarah and {Walkowicz}, Lucianne},
        title = "{The MUSCLES Treasury Survey. II. Intrinsic LY{\ensuremath{\alpha}} and Extreme Ultraviolet Spectra of K and M Dwarfs with Exoplanets*}",
      journal = {\apj},
         year = 2016,
        month = jun,
       volume = {824},
       number = {2},
          eid = {101},
        pages = {101},
          doi = {10.3847/0004-637X/824/2/101},
archivePrefix = {arXiv},
       eprint = {1604.01032},
 primaryClass = {astro-ph.SR},
       adsurl = {https://ui.adsabs.harvard.edu/abs/2016ApJ...824..101Y}
}

@ARTICLE{france_2016_muscles,
       author = {{France}, Kevin and {Loyd}, R.~O. Parke and {Youngblood}, Allison and {Brown}, Alexander and {Schneider}, P. Christian and {Hawley}, Suzanne L. and {Froning}, Cynthia S. and {Linsky}, Jeffrey L. and {Roberge}, Aki and {Buccino}, Andrea P. and {Davenport}, James R.~A. and {Fontenla}, Juan M. and {Kaltenegger}, Lisa and {Kowalski}, Adam F. and {Mauas}, Pablo J.~D. and {Miguel}, Yamila and {Redfield}, Seth and {Rugheimer}, Sarah and {Tian}, Feng and {Vieytes}, Mariela C. and {Walkowicz}, Lucianne M. and {Weisenburger}, Kolby L.},
        title = "{The MUSCLES Treasury Survey. I. Motivation and Overview}",
      journal = {\apj},
         year = 2016,
        month = apr,
       volume = {820},
       number = {2},
          eid = {89},
        pages = {89},
          doi = {10.3847/0004-637X/820/2/89},
archivePrefix = {arXiv},
       eprint = {1602.09142},
 primaryClass = {astro-ph.SR},
       adsurl = {https://ui.adsabs.harvard.edu/abs/2016ApJ...820...89F}
}

@ARTICLE{rajpaul_2016_alfcenb,
       author = {{Rajpaul}, V. and {Aigrain}, S. and {Roberts}, S.},
        title = "{Ghost in the time series: no planet for Alpha Cen B}",
      journal = {\mnras},
         year = 2016,
        month = feb,
       volume = {456},
       number = {1},
        pages = {L6-L10},
          doi = {10.1093/mnrasl/slv164},
archivePrefix = {arXiv},
       eprint = {1510.05598},
 primaryClass = {astro-ph.EP},
       adsurl = {https://ui.adsabs.harvard.edu/abs/2016MNRAS.456L...6R}
}

@ARTICLE{hatzes_2013_alfcenb,
       author = {{Hatzes}, Artie P.},
        title = "{The Radial Velocity Detection of Earth-mass Planets in the Presence of Activity Noise: The Case of {\ensuremath{\alpha}} Centauri Bb}",
      journal = {\apj},
         year = 2013,
        month = jun,
       volume = {770},
       number = {2},
          eid = {133},
        pages = {133},
          doi = {10.1088/0004-637X/770/2/133},
archivePrefix = {arXiv},
       eprint = {1305.4960},
 primaryClass = {astro-ph.SR},
       adsurl = {https://ui.adsabs.harvard.edu/abs/2013ApJ...770..133H}
}

@ARTICLE{sanghi_2025_alfcena,
       author = {{Sanghi}, Aniket and {Beichman}, Charles and {Mawet}, Dimitri and {Balmer}, William O. and {Godoy}, Nicolas and {Pueyo}, Laurent and {Boccaletti}, Anthony and {Sommer}, Max and {Bidot}, Alexis and {Choquet}, Elodie and {Kervella}, Pierre and {Lagage}, Pierre-Olivier and {Leisenring}, Jarron and {Llop-Sayson}, Jorge and {Ressler}, Michael and {Wagner}, Kevin and {Wyatt}, Mark},
        title = "{Worlds Next Door: A Candidate Giant Planet Imaged in the Habitable Zone of {\ensuremath{\alpha}} Centauri A. II. Binary Star Modeling, Planet and Exozodi Search, and Sensitivity Analysis}",
      journal = {\apjl},
         year = 2025,
        month = aug,
       volume = {989},
       number = {2},
          eid = {L23},
        pages = {L23},
          doi = {10.3847/2041-8213/adf53e},
archivePrefix = {arXiv},
       eprint = {2508.03812},
 primaryClass = {astro-ph.EP},
       adsurl = {https://ui.adsabs.harvard.edu/abs/2025ApJ...989L..23S}
}

@ARTICLE{beichman_2025_alfcena,
       author = {{Beichman}, Charles and {Sanghi}, Aniket and {Mawet}, Dimitri and {Kervella}, Pierre and {Wagner}, Kevin and {Quarles}, Billy and {Lissauer}, Jack J. and {Sommer}, Max and {Wyatt}, Mark and {Godoy}, Nicolas and {Balmer}, William O. and {Pueyo}, Laurent and {Llop-Sayson}, Jorge and {Aguilar}, Jonathan and {Akeson}, Rachel and {Belikov}, Ruslan and {Boccaletti}, Anthony and {Choquet}, Elodie and {Fomalont}, Edward and {Henning}, Thomas and {Hines}, Dean and {Hu}, Renyu and {Lagage}, Pierre-Olivier and {Leisenring}, Jarron and {Mang}, James and {Ressler}, Michael and {Serabyn}, Eugene and {Tremblin}, Pascal and {Marie Ygouf} and {Zilinskas}, Mantas},
        title = "{Worlds Next Door: A Candidate Giant Planet Imaged in the Habitable Zone of {\ensuremath{\alpha}} Centauri A. I. Observations, Orbital and Physical Properties, and Exozodi Upper Limits}",
      journal = {\apjl},
         year = 2025,
        month = aug,
       volume = {989},
       number = {2},
          eid = {L22},
        pages = {L22},
          doi = {10.3847/2041-8213/adf53f},
archivePrefix = {arXiv},
       eprint = {2508.03814},
 primaryClass = {astro-ph.EP},
       adsurl = {https://ui.adsabs.harvard.edu/abs/2025ApJ...989L..22B}
}

@ARTICLE{wagner_2021_alfcena,
       author = {{Wagner}, K. and {Boehle}, A. and {Pathak}, P. and {Kasper}, M. and {Arsenault}, R. and {Jakob}, G. and {K{\"a}ufl}, U. and {Leveratto}, S. and {Maire}, A. -L. and {Pantin}, E. and {Siebenmorgen}, R. and {Zins}, G. and {Absil}, O. and {Ageorges}, N. and {Apai}, D. and {Carlotti}, A. and {Choquet}, {\'E}. and {Delacroix}, C. and {Dohlen}, K. and {Duhoux}, P. and {Forsberg}, P. and {Fuenteseca}, E. and {Gutruf}, S. and {Guyon}, O. and {Huby}, E. and {Kampf}, D. and {Karlsson}, M. and {Kervella}, P. and {Kirchbauer}, J. -P. and {Klupar}, P. and {Kolb}, J. and {Mawet}, D. and {N'Diaye}, M. and {Orban de Xivry}, G. and {Quanz}, S.~P. and {Reutlinger}, A. and {Ruane}, G. and {Riquelme}, M. and {Soenke}, C. and {Sterzik}, M. and {Vigan}, A. and {de Zeeuw}, T.},
        title = "{Imaging low-mass planets within the habitable zone of {\ensuremath{\alpha}} Centauri}",
      journal = {Nature Communications},
         year = 2021,
        month = jan,
       volume = {12},
          eid = {922},
        pages = {922},
          doi = {10.1038/s41467-021-21176-6},
archivePrefix = {arXiv},
       eprint = {2102.05159},
 primaryClass = {astro-ph.EP},
       adsurl = {https://ui.adsabs.harvard.edu/abs/2021NatCo..12..922W}
}

@ARTICLE{dumusque_2012_alfcenb,
       author = {{Dumusque}, Xavier and {Pepe}, Francesco and {Lovis}, Christophe and {S{\'e}gransan}, Damien and {Sahlmann}, Johannes and {Benz}, Willy and {Bouchy}, Fran{\c{c}}ois and {Mayor}, Michel and {Queloz}, Didier and {Santos}, Nuno and {Udry}, St{\'e}phane},
        title = "{An Earth-mass planet orbiting {\ensuremath{\alpha}} Centauri B}",
      journal = {\nat},
         year = 2012,
        month = nov,
       volume = {491},
       number = {7423},
        pages = {207-211},
          doi = {10.1038/nature11572},
       adsurl = {https://ui.adsabs.harvard.edu/abs/2012Natur.491..207D}
}

@ARTICLE{lustig-yaeger_transmission_2023,
       author = {{Lustig-Yaeger}, Jacob and {Fu}, Guangwei and {May}, E.~M. and {Ceballos}, Kevin N. Ortiz and {Moran}, Sarah E. and {Peacock}, Sarah and {Stevenson}, Kevin B. and {Kirk}, James and {L{\'o}pez-Morales}, Mercedes and {MacDonald}, Ryan J. and {Mayorga}, L.~C. and {Sing}, David K. and {Sotzen}, Kristin S. and {Valenti}, Jeff A. and {Redai}, J{\'e}a I. Adams and {Alam}, Munazza K. and {Batalha}, Natasha E. and {Bennett}, Katherine A. and {Gonzalez-Quiles}, Junellie and {Kruse}, Ethan and {Lothringer}, Joshua D. and {Rustamkulov}, Zafar and {Wakeford}, Hannah R.},
        title = "{A JWST transmission spectrum of the nearby Earth-sized exoplanet LHS 475 b}",
      journal = {Nature Astronomy},
         year = 2023,
        month = nov,
       volume = {7},
        pages = {1317-1328},
          doi = {10.1038/s41550-023-02064-z},
archivePrefix = {arXiv},
       eprint = {2301.04191},
 primaryClass = {astro-ph.EP},
       adsurl = {https://ui.adsabs.harvard.edu/abs/2023NatAs...7.1317L}
}

@ARTICLE{rustamkulov_analysis_2022,
       author = {{Rustamkulov}, Zafar and {Sing}, David K. and {Liu}, Rongrong and {Wang}, Ashley},
        title = "{Analysis of a JWST NIRSpec Lab Time Series: Characterizing Systematics, Recovering Exoplanet Transit Spectroscopy, and Constraining a Noise Floor}",
      journal = {\apjl},
         year = 2022,
        month = mar,
       volume = {928},
       number = {1},
          eid = {L7},
        pages = {L7},
          doi = {10.3847/2041-8213/ac5b6f},
archivePrefix = {arXiv},
       eprint = {2203.04173},
 primaryClass = {astro-ph.EP},
       adsurl = {https://ui.adsabs.harvard.edu/abs/2022ApJ...928L...7R}
}

@ARTICLE{pagano_2004_stis,
       author = {{Pagano}, I. and {Linsky}, J.~L. and {Valenti}, J. and {Duncan}, D.~K.},
        title = "{HST/STIS high resolution echelle spectra of <ASTROBJ>{\ensuremath{\alpha}} Centauri A</ASTROBJ> (G2 V)}",
      journal = {\aap},
         year = 2004,
        month = feb,
       volume = {415},
        pages = {331-348},
          doi = {10.1051/0004-6361:20034002},
archivePrefix = {arXiv},
       eprint = {astro-ph/0310901},
 primaryClass = {astro-ph},
       adsurl = {https://ui.adsabs.harvard.edu/abs/2004A&A...415..331P}
}

@ARTICLE{henry_1996_caii,
       author = {{Henry}, Todd J. and {Soderblom}, David R. and {Donahue}, Robert A. and {Baliunas}, Sallie L.},
        title = "{A Survey of Ca II H and K Chromospheric Emission in Southern Solar-Type Stars}",
      journal = {\aj},
         year = 1996,
        month = jan,
       volume = {111},
        pages = {439},
          doi = {10.1086/117796},
       adsurl = {https://ui.adsabs.harvard.edu/abs/1996AJ....111..439H}
}

@ARTICLE{saar_osten_1997,
       author = {{Saar}, S.~H. and {Osten}, R.~A.},
        title = "{Rotation, turbulence and evidence for magnetic fields in southern dwarfs}",
      journal = {\mnras},
         year = 1997,
        month = feb,
       volume = {284},
       number = {4},
        pages = {803-810},
          doi = {10.1093/mnras/284.4.803},
       adsurl = {https://ui.adsabs.harvard.edu/abs/1997MNRAS.284..803S}
}

@ARTICLE{johnstone_2021_xuv,
       author = {{Johnstone}, C.~P. and {Bartel}, M. and {G{\"u}del}, M.},
        title = "{The active lives of stars: A complete description of the rotation and XUV evolution of F, G, K, and M dwarfs}",
      journal = {\aap},
         year = 2021,
        month = may,
       volume = {649},
          eid = {A96},
        pages = {A96},
          doi = {10.1051/0004-6361/202038407},
archivePrefix = {arXiv},
       eprint = {2009.07695},
 primaryClass = {astro-ph.SR},
       adsurl = {https://ui.adsabs.harvard.edu/abs/2021A&A...649A..96J}
}

@ARTICLE{bowyer_1991_euve,
       author = {{Bowyer}, S. and {Malina}, R.~F.},
        title = "{The extreme ultraviolet explorer mission}",
      journal = {Advances in Space Research},
         year = 1991,
        month = jan,
       volume = {11},
       number = {11},
        pages = {205-215},
          doi = {10.1016/0273-1177(91)90077-W},
       adsurl = {https://ui.adsabs.harvard.edu/abs/1991AdSpR..11k.205B}
}

@BOOK{mariska_1992_transition,
       author = {{Mariska}, John T.},
        title = "{The Solar Transition Region}",
         year = 1992,
       adsurl = {https://ui.adsabs.harvard.edu/abs/1992str..book.....M}
}

@INPROCEEDINGS{arney_2025_hwo,
       author = {{Arney}, Giada},
        title = "{Are We Alone? The Search for Life on Habitable Worlds}",
    booktitle = {American Astronomical Society Meeting Abstracts \#245},
         year = 2025,
       series = {American Astronomical Society Meeting Abstracts},
       volume = {245},
        month = jan,
          eid = {331.01},
        pages = {331.01},
       adsurl = {https://ui.adsabs.harvard.edu/abs/2025AAS...24533101A}
}

@ARTICLE{binder_2024_hwo,
       author = {{Binder}, Breanna A. and {Peacock}, Sarah and {Schwieterman}, Edward W. and {Turnbull}, Margaret C. and {Virgen}, Azariel Y. and {Kane}, Stephen R. and {Farrish}, Alison and {Garcia-Sage}, Katherine},
        title = "{X-Ray Emission of Nearby Low-mass and Sunlike Stars with Directly Imageable Habitable Zones}",
      journal = {\apjs},
         year = 2024,
        month = nov,
       volume = {275},
       number = {1},
          eid = {1},
        pages = {1},
          doi = {10.3847/1538-4365/ad71d6},
archivePrefix = {arXiv},
       eprint = {2407.21247},
 primaryClass = {astro-ph.HE},
       adsurl = {https://ui.adsabs.harvard.edu/abs/2024ApJS..275....1B}
}

@ARTICLE{thebault_2025_binaries,
       author = {{Thebault}, P. and {Bonanni}, D.},
        title = "{A complete census of planet-hosting binaries}",
      journal = {\aap},
         year = 2025,
        month = aug,
       volume = {700},
          eid = {A106},
        pages = {A106},
          doi = {10.1051/0004-6361/202555457},
archivePrefix = {arXiv},
       eprint = {2506.18759},
 primaryClass = {astro-ph.EP},
       adsurl = {https://ui.adsabs.harvard.edu/abs/2025A&A...700A.106T}
}

@ARTICLE{tian_2015_atmospheric_escape,
       author = {{Tian}, Feng},
        title = "{Atmospheric Escape from Solar System Terrestrial Planets and Exoplanets}",
      journal = {Annual Review of Earth and Planetary Sciences},
         year = 2015,
        month = may,
       volume = {43},
        pages = {459-476},
          doi = {10.1146/annurev-earth-060313-054834},
       adsurl = {https://ui.adsabs.harvard.edu/abs/2015AREPS..43..459T}
}

@ARTICLE{warren_1998_dem,
       author = {{Warren}, H.~P. and {Mariska}, J.~T. and {Lean}, J.},
        title = "{A new reference spectrum for the EUV irradiance of the quiet Sun 1. Emission measure formulation}",
      journal = {\jgr},
         year = 1998,
        month = jun,
       volume = {103},
       number = {A6},
        pages = {12077-12090},
          doi = {10.1029/98JA00810},
       adsurl = {https://ui.adsabs.harvard.edu/abs/1998JGR...10312077W}
}

@ARTICLE{duvvuri_2025_euv,
       author = {{Duvvuri}, Girish M. and {Berta-Thompson}, Zachory K. and {Pineda}, J. Sebastian and {France}, Kevin and {Brown}, Alexander and {Youngblood}, Allison and {Wilson}, David J. and {Froning}, Cynthia S. and {Schneider}, P. Christian and {Ayres}, Thomas and et al.},
        title = "{Stellar Library of Differential Emission Measures and Extreme Ultraviolet Spectra: Dwarf Stars Observed by the Extreme Ultraviolet Explorer}",
      journal = {\apj},
         year = 2025,
        month = nov,
       volume = {993},
       number = {1},
          eid = {138},
        pages = {138},
          doi = {10.3847/1538-4357/ae06a7},
       adsurl = {https://ui.adsabs.harvard.edu/abs/2025ApJ...993..138D}
}

@ARTICLE{chadney_2015_mass-loss,
       author = {{Chadney}, J.~M. and {Galand}, M. and {Unruh}, Y.~C. and {Koskinen}, T.~T. and {Sanz-Forcada}, J.},
        title = "{XUV-driven mass loss from extrasolar giant planets orbiting active stars}",
      journal = {\icarus},
         year = 2015,
        month = apr,
       volume = {250},
        pages = {357-367},
          doi = {10.1016/j.icarus.2014.12.012},
archivePrefix = {arXiv},
       eprint = {1412.3380},
 primaryClass = {astro-ph.SR},
       adsurl = {https://ui.adsabs.harvard.edu/abs/2015Icar..250..357C}
}

@ARTICLE{hartkopf_2001_orb6,
       author = {{Hartkopf}, William I. and {Mason}, Brian D. and {Worley}, Charles E.},
        title = "{The 2001 US Naval Observatory Double Star CD-ROM. II. The Fifth Catalog of Orbits of Visual Binary Stars}",
      journal = {\aj},
         year = 2001,
        month = dec,
       volume = {122},
       number = {6},
        pages = {3472-3479},
          doi = {10.1086/323921},
       adsurl = {https://ui.adsabs.harvard.edu/abs/2001AJ....122.3472H}
}

@ARTICLE{eggenberger_2008_70oph,
       author = {{Eggenberger}, P. and {Miglio}, A. and {Carrier}, F. and {Fernandes}, J. and {Santos}, N.~C.},
        title = "{Analysis of 70 Ophiuchi AB including seismic constraints}",
      journal = {\aap},
         year = 2008,
        month = may,
       volume = {482},
       number = {2},
        pages = {631-638},
          doi = {10.1051/0004-6361:20078624},
archivePrefix = {arXiv},
       eprint = {0802.3576},
 primaryClass = {astro-ph},
       adsurl = {https://ui.adsabs.harvard.edu/abs/2008A&A...482..631E}
}

@ARTICLE{piccotti_2020_binaries,
       author = {{Piccotti}, Luca and {Docobo}, Jos{\'e} {\'A}ngel and {Carini}, Roberta and {Tamazian}, Vakhtang S. and {Brocato}, Enzo and {Andrade}, Manuel and {Campo}, Pedro P.},
        title = "{A study of the physical properties of SB2s with both the visual and spectroscopic orbits}",
      journal = {\mnras},
         year = 2020,
        month = feb,
       volume = {492},
       number = {2},
        pages = {2709-2721},
          doi = {10.1093/mnras/stz3616},
       adsurl = {https://ui.adsabs.harvard.edu/abs/2020MNRAS.492.2709P}
}

@ARTICLE{takovinin_2017_triple,
       author = {{Tokovinin}, Andrei},
        title = "{Orbit Alignment in Triple Stars}",
      journal = {\apj},
         year = 2017,
        month = aug,
       volume = {844},
       number = {2},
          eid = {103},
        pages = {103},
          doi = {10.3847/1538-4357/aa7746},
archivePrefix = {arXiv},
       eprint = {1706.00748},
 primaryClass = {astro-ph.SR},
       adsurl = {https://ui.adsabs.harvard.edu/abs/2017ApJ...844..103T}
}

@ARTICLE{luck_2017_abundances,
       author = {{Luck}, R. Earle},
        title = "{Abundances in the Local Region II: F, G, and K Dwarfs and Subgiants}",
      journal = {\aj},
         year = 2017,
        month = jan,
       volume = {153},
       number = {1},
          eid = {21},
        pages = {21},
          doi = {10.3847/1538-3881/153/1/21},
archivePrefix = {arXiv},
       eprint = {1611.02897},
 primaryClass = {astro-ph.SR},
       adsurl = {https://ui.adsabs.harvard.edu/abs/2017AJ....153...21L}
}

@ARTICLE{luck_2018_abundances,
       author = {{Luck}, R. Earle},
        title = "{Abundances in the Local Region. III. Southern F, G, and K Dwarfs}",
      journal = {\aj},
         year = 2018,
        month = mar,
       volume = {155},
       number = {3},
          eid = {111},
        pages = {111},
          doi = {10.3847/1538-3881/aaa9b5},
       adsurl = {https://ui.adsabs.harvard.edu/abs/2018AJ....155..111L}
}

@ARTICLE{malkov_2012_binaries,
       author = {{Malkov}, O. Yu. and {Tamazian}, V.~S. and {Docobo}, J.~A. and {Chulkov}, D.~A.},
        title = "{Dynamical masses of a selected sample of orbital binaries}",
      journal = {\aap},
         year = 2012,
        month = oct,
       volume = {546},
          eid = {A69},
        pages = {A69},
          doi = {10.1051/0004-6361/201219774},
       adsurl = {https://ui.adsabs.harvard.edu/abs/2012A&A...546A..69M}
}

@ARTICLE{soubiran_2024_gaia,
       author = {{Soubiran}, C. and {Creevey}, O.~L. and {Lagarde}, N. and {Brouillet}, N. and {Jofr{\'e}}, P. and {Casamiquela}, L. and {Heiter}, U. and {Aguilera-G{\'o}mez}, C. and {Vitali}, S. and {Worley}, C. and {de Brito Silva}, D.},
        title = "{Gaia FGK benchmark stars: Fundamental T$_{eff}$ and log g of the third version}",
      journal = {\aap},
         year = 2024,
        month = feb,
       volume = {682},
          eid = {A145},
        pages = {A145},
          doi = {10.1051/0004-6361/202347136},
archivePrefix = {arXiv},
       eprint = {2310.11302},
 primaryClass = {astro-ph.SR},
       adsurl = {https://ui.adsabs.harvard.edu/abs/2024A&A...682A.145S}
}

@ARTICLE{wilson_2021_trappist,
       author = {{Wilson}, David J. and {Froning}, Cynthia S. and {Duvvuri}, Girish M. and {France}, Kevin and {Youngblood}, Allison and {Schneider}, P. Christian and {Berta-Thompson}, Zachory and {Brown}, Alexander and {Buccino}, Andrea P. and {Hawley}, Suzanne and et al.},
        title = "{The Mega-MUSCLES Spectral Energy Distribution of TRAPPIST-1}",
      journal = {\apj},
         year = 2021,
        month = apr,
       volume = {911},
       number = {1},
          eid = {18},
        pages = {18},
          doi = {10.3847/1538-4357/abe771},
archivePrefix = {arXiv},
       eprint = {2102.11415},
 primaryClass = {astro-ph.SR},
       adsurl = {https://ui.adsabs.harvard.edu/abs/2021ApJ...911...18W}
}

@INPROCEEDINGS{allard_2016_phoenix,
       author = {{Allard}, F.},
        title = "{The PHOENIX Model Atmosphere Grid for Stars}",
    booktitle = {SF2A-2016: Proceedings of the Annual meeting of the French Society of Astronomy and Astrophysics},
         year = 2016,
       editor = {{Reyl{\'e}}, C. and {Richard}, J. and {Cambr{\'e}sy}, L. and {Deleuil}, M. and {P{\'e}contal}, E. and {Tresse}, L. and {Vauglin}, I.},
        month = dec,
        pages = {223-227},
       adsurl = {https://ui.adsabs.harvard.edu/abs/2016sf2a.conf..223A}
}

@ARTICLE{baraffe_2015_model,
       author = {{Baraffe}, Isabelle and {Homeier}, Derek and {Allard}, France and {Chabrier}, Gilles},
        title = "{New evolutionary models for pre-main sequence and main sequence low-mass stars down to the hydrogen-burning limit}",
      journal = {\aap},
         year = 2015,
        month = may,
       volume = {577},
          eid = {A42},
        pages = {A42},
          doi = {10.1051/0004-6361/201425481},
archivePrefix = {arXiv},
       eprint = {1503.04107},
 primaryClass = {astro-ph.SR},
       adsurl = {https://ui.adsabs.harvard.edu/abs/2015A&A...577A..42B}
}

@ARTICLE{sanz-forcada_2011_dem,
       author = {{Sanz-Forcada}, J. and {Micela}, G. and {Ribas}, I. and {Pollock}, A.~M.~T. and {Eiroa}, C. and {Velasco}, A. and {Solano}, E. and {Garc{\'\i}a-{\'A}lvarez}, D.},
        title = "{Estimation of the XUV radiation onto close planets and their evaporation}",
      journal = {\aap},
         year = 2011,
        month = aug,
       volume = {532},
          eid = {A6},
        pages = {A6},
          doi = {10.1051/0004-6361/201116594},
archivePrefix = {arXiv},
       eprint = {1105.0550},
 primaryClass = {astro-ph.EP},
       adsurl = {https://ui.adsabs.harvard.edu/abs/2011A&A...532A...6S}
}

@ARTICLE{tian_2008_xuv,
       author = {{Tian}, Feng and {Kasting}, James F. and {Liu}, Han-Li and {Roble}, Raymond G.},
        title = "{Hydrodynamic planetary thermosphere model: 1. Response of the Earth's thermosphere to extreme solar EUV conditions and the significance of adiabatic cooling}",
      journal = {Journal of Geophysical Research (Planets)},
         year = 2008,
        month = may,
       volume = {113},
       number = {E5},
          eid = {E05008},
        pages = {E05008},
          doi = {10.1029/2007JE002946},
       adsurl = {https://ui.adsabs.harvard.edu/abs/2008JGRE..113.5008T}
}

@ARTICLE{nell_2024_sistine,
       author = {{Nell}, Nicholas and {France}, Kevin and {Kruczek}, Nicholas and {Fleming}, Brian and {Ulrich}, Stefan and {Behr}, Patrick and {Quijada}, Manuel A. and {Hoyo}, Javier Del and {Hennessy}, John},
        title = "{The assembly, characterization, and performance of SISTINE}",
      journal = {Journal of Astronomical Telescopes, Instruments, and Systems},
         year = 2024,
        month = jul,
       volume = {10},
          eid = {035003},
        pages = {035003},
          doi = {10.1117/1.JATIS.10.3.035003},
archivePrefix = {arXiv},
       eprint = {2410.02893},
 primaryClass = {astro-ph.IM},
       adsurl = {https://ui.adsabs.harvard.edu/abs/2024JATIS..10c5003N}
}

@article{woods_solar_2009,
	title = {Solar {Irradiance} {Reference} {Spectra} ({SIRS}) for the 2008 {Whole} {Heliosphere} {Interval} ({WHI})},
	volume = {36},
	copyright = {Copyright 2009 by the American Geophysical Union.},
	issn = {1944-8007},
	url = {https://onlinelibrary.wiley.com/doi/abs/10.1029/2008GL036373},
	doi = {10.1029/2008GL036373},
	language = {en},
	number = {1},
	urldate = {2024-06-28},
	journal = {Geophysical Research Letters},
	author = {Woods, Thomas N. and Chamberlin, Phillip C. and Harder, Jerald W. and Hock, Rachel A. and Snow, Martin and Eparvier, Francis G. and Fontenla, Juan and McClintock, William E. and Richard, Erik C.},
	year = {2009},
	note = {\_eprint: https://onlinelibrary.wiley.com/doi/pdf/10.1029/2008GL036373},
}

@misc{foreman-mackey_emcee_2013,
	title = {emcee: {The} {MCMC} {Hammer}},
	shorttitle = {emcee},
	url = {http://arxiv.org/abs/1202.3665},
	doi = {10.1086/670067},
	language = {en},
	urldate = {2024-08-21},
	author = {Foreman-Mackey, Daniel and Hogg, David W. and Lang, Dustin and Goodman, Jonathan},
	month = nov,
	year = {2013},
	note = {arXiv:1202.3665 [astro-ph, physics:physics, stat]},
}

@article{sing_hubble_2019,
	title = {The {Hubble} {Space} {Telescope} {PanCET} {Program}: {Exospheric} {Mg} ii and {Fe} ii in the {Near}-ultraviolet {Transmission} {Spectrum} of {WASP}-121b {Using} {Jitter} {Decorrelation}},
	volume = {158},
	issn = {1538-3881},
	shorttitle = {The {Hubble} {Space} {Telescope} {PanCET} {Program}},
	url = {https://dx.doi.org/10.3847/1538-3881/ab2986},
	doi = {10.3847/1538-3881/ab2986},
	language = {en},
	number = {2},
	urldate = {2024-09-12},
	journal = {The Astronomical Journal},
	author = {Sing, David K. and Lavvas, Panayotis and Ballester, Gilda E. and Etangs, Alain Lecavelier des and Marley, Mark S. and Nikolov, Nikolay and Ben-Jaffel, Lotfi and Bourrier, Vincent and Buchhave, Lars A. and Deming, Drake L. and Ehrenreich, David and Mikal-Evans, Thomas and Kataria, Tiffany and Lewis, Nikole K. and López-Morales, Mercedes and Muñoz, Antonio García and Henry, Gregory W. and Sanz-Forcada, Jorge and Spake, Jessica J. and Wakeford, Hannah R. and collaboration), (The PanCET},
	month = aug,
	year = {2019},
	note = {Publisher: The American Astronomical Society},
	pages = {91},
}

@ARTICLE{wright_2011_activity,
       author = {{Wright}, Nicholas J. and {Drake}, Jeremy J. and {Mamajek}, Eric E. and {Henry}, Gregory W.},
        title = "{The Stellar-activity-Rotation Relationship and the Evolution of Stellar Dynamos}",
      journal = {\apj},
         year = 2011,
        month = dec,
       volume = {743},
       number = {1},
          eid = {48},
        pages = {48},
          doi = {10.1088/0004-637X/743/1/48},
archivePrefix = {arXiv},
       eprint = {1109.4634},
 primaryClass = {astro-ph.SR},
       adsurl = {https://ui.adsabs.harvard.edu/abs/2011ApJ...743...48W}
}

@article{behr_muscles_2023,
	title = {The {MUSCLES} {Extension} for {Atmospheric} {Transmission} {Spectroscopy}: {UV} and {X}-{Ray} {Host}-star {Observations} for {JWST} {ERS} \& {GTO} {Targets}},
	volume = {166},
	issn = {1538-3881},
	shorttitle = {The {MUSCLES} {Extension} for {Atmospheric} {Transmission} {Spectroscopy}},
	url = {https://dx.doi.org/10.3847/1538-3881/acdb70},
	doi = {10.3847/1538-3881/acdb70},
	language = {en},
	number = {1},
	urldate = {2024-10-02},
	journal = {The Astronomical Journal},
	author = {Behr, Patrick R. and France, Kevin and Brown, Alexander and Duvvuri, Girish and Bean, Jacob L. and Berta-Thompson, Zachory and Froning, Cynthia and Miguel, Yamila and Pineda, J. Sebastian and Wilson, David J. and Youngblood, Allison},
	month = jun,
	year = {2023},
	note = {Publisher: The American Astronomical Society},
	pages = {35},
}

@article{schwieterman_overview_2024,
	title = {An {Overview} of {Exoplanet} {Biosignatures}},
	volume = {90},
	issn = {1529-6466},
	url = {https://doi.org/10.2138/rmg.2024.90.13},
	doi = {10.2138/rmg.2024.90.13},
	number = {1},
	urldate = {2024-10-03},
	journal = {Reviews in Mineralogy and Geochemistry},
	author = {Schwieterman, Edward W. and Leung, Michaela},
	month = jul,
	year = {2024},
	pages = {465--514},
}

@article{duvvuri_reconstructing_2021,
	title = {Reconstructing the {Extreme} {Ultraviolet} {Emission} of {Cool} {Dwarfs} {Using} {Differential} {Emission} {Measure} {Polynomials}},
	volume = {913},
	issn = {0004-637X},
	url = {https://dx.doi.org/10.3847/1538-4357/abeaaf},
	doi = {10.3847/1538-4357/abeaaf},
	language = {en},
	number = {1},
	urldate = {2024-10-04},
	journal = {The Astrophysical Journal},
	author = {Duvvuri, Girish M. and Pineda, J. Sebastian and Berta-Thompson, Zachory K. and Brown, Alexander and France, Kevin and Kowalski, Adam F. and Redfield, Seth and Tilipman, Dennis and Vieytes, Mariela C. and Wilson, David J. and Youngblood, Allison and Froning, Cynthia S. and Linsky, Jeffrey and Loyd, R. O. Parke and Mauas, Pablo and Miguel, Yamila and Newton, Elisabeth R. and Rugheimer, Sarah and Schneider, P. Christian},
	month = may,
	year = {2021},
	note = {Publisher: The American Astronomical Society},
	pages = {40},
}

@article{aguirre_radiation_2023,
	title = {The {Radiation} {Environments} of {Middle}-aged {F}-type {Stars}},
	volume = {956},
	issn = {0004-637X},
	url = {https://dx.doi.org/10.3847/1538-4357/aced9f},
	doi = {10.3847/1538-4357/aced9f},
	language = {en},
	number = {2},
	urldate = {2024-10-21},
	journal = {The Astrophysical Journal},
	author = {Cruz Aguirre, F. and France, K. and Nell, N. and Kruczek, N. and Fleming, B. and Hinton, P. C. and Ulrich, S. and Behr, P. R.},
	month = oct,
	year = {2023},
	note = {Publisher: The American Astronomical Society},
	pages = {79},
}

@ARTICLE{lammer_2003_escape,
       author = {{Lammer}, H. and {Selsis}, F. and {Ribas}, I. and {Guinan}, E.~F. and {Bauer}, S.~J. and {Weiss}, W.~W.},
        title = "{Atmospheric Loss of Exoplanets Resulting from Stellar X-Ray and Extreme-Ultraviolet Heating}",
      journal = {\apjl},
         year = 2003,
        month = dec,
       volume = {598},
       number = {2},
        pages = {L121-L124},
          doi = {10.1086/380815},
       adsurl = {https://ui.adsabs.harvard.edu/abs/2003ApJ...598L.121L}
}

@article{kaltenegger_calculating_2013,
	title = {{CALCULATING} {THE} {HABITABLE} {ZONE} {OF} {BINARY} {STAR} {SYSTEMS}. {I}. {S}-{TYPE} {BINARIES}},
	volume = {777},
	issn = {0004-637X},
	url = {https://dx.doi.org/10.1088/0004-637X/777/2/165},
	doi = {10.1088/0004-637X/777/2/165},
	language = {en},
	number = {2},
	urldate = {2024-10-24},
	journal = {The Astrophysical Journal},
	author = {Kaltenegger, Lisa and Haghighipour, Nader},
	month = oct,
	year = {2013},
	note = {Publisher: The American Astronomical Society},
	pages = {165},
}

@article{akeson_precision_2021,
	title = {Precision {Millimeter} {Astrometry} of the α {Centauri} {AB} {System}},
	volume = {162},
	issn = {1538-3881},
	url = {https://dx.doi.org/10.3847/1538-3881/abfaff},
	doi = {10.3847/1538-3881/abfaff},
	language = {en},
	number = {1},
	urldate = {2024-10-24},
	journal = {The Astronomical Journal},
	author = {Akeson, Rachel and Beichman, Charles and Kervella, Pierre and Fomalont, Edward and Benedict, G. Fritz},
	month = jun,
	year = {2021},
	note = {Publisher: The American Astronomical Society},
	pages = {14},
}

@article{ayres_cycles_2023,
	title = {The {Cycles} of {Alpha} {Centauri}: {Double} {Dipping} of {AB}},
	volume = {166},
	issn = {1538-3881},
	shorttitle = {The {Cycles} of {Alpha} {Centauri}},
	url = {https://dx.doi.org/10.3847/1538-3881/acfef5},
	doi = {10.3847/1538-3881/acfef5},
	language = {en},
	number = {5},
	urldate = {2024-10-24},
	journal = {The Astronomical Journal},
	author = {Ayres, Thomas},
	month = nov,
	year = {2023},
	note = {Publisher: The American Astronomical Society},
	pages = {212},
}

@article{meadows_exoplanet_2018,
	title = {Exoplanet {Biosignatures}: {Understanding} {Oxygen} as a {Biosignature} in the {Context} of {Its} {Environment}},
	volume = {18},
	issn = {1531-1074},
	shorttitle = {Exoplanet {Biosignatures}},
	url = {https://www.liebertpub.com/doi/10.1089/ast.2017.1727},
	doi = {10.1089/ast.2017.1727},
	number = {6},
	urldate = {2024-10-24},
	journal = {Astrobiology},
	author = {Meadows, Victoria S. and Reinhard, Christopher T. and Arney, Giada N. and Parenteau, Mary N. and Schwieterman, Edward W. and Domagal-Goldman, Shawn D. and Lincowski, Andrew P. and Stapelfeldt, Karl R. and Rauer, Heike and DasSarma, Shiladitya and Hegde, Siddharth and Narita, Norio and Deitrick, Russell and Lustig-Yaeger, Jacob and Lyons, Timothy W. and Siegler, Nicholas and Grenfell, J. Lee},
	month = jun,
	year = {2018},
	note = {Publisher: Mary Ann Liebert, Inc., publishers},
	pages = {630--662},
}

@article{morel_chemical_2018,
	title = {The chemical composition of α {Centauri} {AB} revisited},
	volume = {615},
	copyright = {© ESO 2018},
	issn = {0004-6361, 1432-0746},
	url = {https://www.aanda.org/articles/aa/abs/2018/07/aa33125-18/aa33125-18.html},
	doi = {10.1051/0004-6361/201833125},
	language = {en},
	urldate = {2024-10-24},
	journal = {Astronomy \& Astrophysics},
	author = {Morel, Thierry},
	month = jul,
	year = {2018},
	note = {Publisher: EDP Sciences},
	pages = {A172},
}
\bibliographystyle{aasjournal}

\end{document}